\documentclass[sigplan,10pt,twocolumn,screen]{acmart}

\acmSubmissionID{332}

\usepackage{amsmath}

\usepackage{enumitem}
\usepackage{multirow}
\usepackage{xspace}

\usepackage{amssymb}
\usepackage{pifont}
\usepackage{breakurl}

\usepackage[ruled,vlined,linesnumbered]{algorithm2e}
\usepackage{graphicx}
\usepackage{float}

\usepackage{hyperref}

\newcommand{\tool}{\textsc{LLMSbt}\xspace}

\newcommand{\F}{Fig.}

\newcommand{\T}{Table}
\renewcommand{\S}{Sec.}
\newcommand{\A}{Alg.}

\usepackage[normalem]{ulem}

\newcommand*\circled[1]{\ding{\numexpr171+#1\relax}}

\newcommand{\mysubref}[2]{\hyperref[#1]{\ref*{#1}(#2)}}

\newcommand{\parh}[1]{\noindent\textbf{#1}}

\newcommand{\uparh}[1]{\noindent\underline{#1}}
\newcommand{\sparh}[1]{\noindent\underline{\textbf{#1}}}

\begin{document}

\acmYear{2026}\copyrightyear{2026}
\setcopyright{cc}
\setcctype[4.0]{by-nc-nd}
\acmConference[EUROSYS '26]{European Conference on Computer Systems}{April 27--30, 2026}{Edinburgh, Scotland Uk}
\acmBooktitle{European Conference on Computer Systems (EUROSYS '26), April 27--30, 2026, Edinburgh, Scotland Uk}
\acmDOI{10.1145/3767295.3803600}
\acmISBN{979-8-4007-2212-7/26/04}

\date{}

\title[]{No More Translation at Runtime: LLM-Empowered Static Binary Translation}

\author{Zhibo Liu}
\affiliation{\institution{State Key Laboratory of Novel Software Technology, \\ Nanjing University}
\city{Nanjing}
  \country{China}
  }
\email{zhiboliu@nju.edu.cn}
\authornotemark[1]

\author{Huaijin Wang}
\affiliation{\institution{The Hong Kong University of \\ Science and Technology}
\city{Hong Kong}
  \country{China}
  }
\email{hwangdz@cse.ust.hk}

\author{Wai Kin Wong}
\affiliation{\institution{The Hong Kong University of \\ Science and Technology}
\city{Hong Kong}
  \country{China}
  }
\email{wkwongal@cse.ust.hk}

\author{Daoyuan Wu}
\affiliation{\institution{Lingnan University}
\city{Hong Kong}
  \country{China}
  }
\email{daoyuanwu@ln.edu.hk}
\authornote{Work done while the author was associated with HKUST.}

\author{Shuai Wang}
\authornote{Corresponding author.}
\affiliation{\institution{The Hong Kong University of \\ Science and Technology}
\city{Hong Kong}
  \country{China}
  }
\email{shuaiw@cse.ust.hk}

\renewcommand{\shortauthors}{Zhibo Liu et al.}
\renewcommand{\authors}{Zhibo Liu, Huaijin Wang, Wai Kin Wong, Daoyuan Wu, Shuai Wang}

\begin{abstract}
While AArch64 CPUs are becoming strong market contenders, their software
ecosystem lags behind the mature x86-64 environment, hindering the adoption of
the new architectures and impacting user experience. 
\textit{Binary translation} bridges this divide by converting binary code
from one architecture (e.g., x86-64) to run on
another (e.g., AArch64), allowing legacy software to benefit from modern
hardware's performance and energy efficiency advantages.

Current translation methods are typically either dynamic,
which adds significant runtime overhead, or static, 
which struggles with reliability due to the inherent complexities of binary
analysis. 
This paper introduces a new static, assembly-to-assembly 
translation paradigm that transforms binary code ahead of execution, 
generating portable, efficient native-like binaries that run on AArch64
devices without runtime frameworks. 
Benefiting from recent breakthroughs in large language models (LLMs), we provide
a practical and automated translation engine that produces high-quality code
with minimal human intervention. To ensure correctness, we introduce a
crucial verification step, where we split the assembly code into simplified
snippets, enabling efficient and scalable semantic verification.

Our evaluation shows that this approach significantly outperforms existing
open-source solutions with a large margin, producing binaries with
near-native performance. Furthermore, it shows
substantial improvements over 
the leading industrial translator, ExaGear, 
illuminating a promising new direction for cross-architecture
binary translation research.

\end{abstract}



\maketitle

\section{Introduction}
\label{sec:introduction}

The computing hardware landscape is shifting away from the long-standing
dominance of x86-64, driven by the rise of alternative Instruction
Set Architectures (ISAs) like ARM and
RISC-V~\cite{frumusanu2020amazon,shah2021we}. 
AArch64 has already achieved significant success in mobile devices and is now
making remarkable strides in the PC market with processors like Apple's Silicon
chips~\cite{apple-silicon,apple-m4}. 
Their high performance and energy efficiency have spurred  
major manufacturers to develop and promote AArch64 CPUs.
Notable examples include Huawei's Kunpeng server chips~\cite{hisilicon} and
Qualcomm's Snapdragon X Elite processor for Windows
laptops~\cite{qualcomm-snapdragon,windows-prism}.

Despite the advantages of ARM-based chips, their widespread adoption is
hindered by a major software hurdle. While the AArch64 software ecosystem is
thriving, decades of software development have been optimized for x86-64. Legacy
applications often lack available source code or complete build environments,
making a source-based recompilation non-viable~\cite{apple-porting}. 

To bridge this gap, chip manufacturers and researchers have turned to binary
translation\cite{bellard2005qemu,apple-rosetta,windows-prism}, which
converts binary code from one ISA to another, allowing seamless
software migration from an old architecture to a new one. 
Prominent examples include Apple's Rosetta 2~\cite{apple-rosetta} for macOS,
Huawei's ExaGear~\cite{exagear} for its Kunpeng-powered Linux systems, 
and Microsoft's
Prism~\cite{windows-prism} for Windows on ARM.

While existing binary translation solutions have facilitated the rapid growth of
the AArch64 ecosystem, 
they are often limited by their low reliability or high runtime overhead.
Specifically, static translators lift assembly code to a higher-level,
platform-independent intermediate representation (IR) (e.g., LLVM IR) before
recompiling it for a new platform~\cite{yadavalli2019raising,retdec}. 
This process is fundamentally challenged by the undecidable problem of
statically recovering high-level semantic
information~\cite{horspool1980approach}, such as symbols and variables, that is
discarded during compilation.
Differently, dynamic solutions translate code at
runtime~\cite{engelke2020instrew,bartholomew2006qemu,exagear}, which not only
incurs extra performance overhead, but also  
restricts the portability of translated software, preventing it from migrating to
platforms without the translation framework.

Moreover, state-of-the-art binary translation tools are closed-source and
platform or CPU-specific~\cite{apple-rosetta,exagear,windows-prism}, creating barriers
for research aiming to enhance modern binary translation
techniques. The complexity of modern ISAs 
makes developing a cross-architecture translator a daunting task, requiring
substantial expertise and effort to identify equivalent instructions across ISAs
and examine their
subtle semantic differences. 
These challenges have slowed academic research,
highlighting the need for an automated solution.

To address these limitations, we present a new paradigm for
cross-architecture binary translation. 
Leveraging recent advances in reverse engineering research and LLMs, we propose
a \textit{static assembly-to-assembly} translation paradigm that
converts x86-64 binaries to AArch64 \textit{ahead of time}, eliminating runtime
translation overhead while producing portable binaries that can run natively on
AArch64 platforms. Furthermore, given LLMs' success in multilingual translation
and various coding
tasks~\cite{chang2024survey,lee2023guess,cummins2024meta,guo2024deepseek,10.1145/3597503.3639091},
we explore their potential to facilitate binary translator development,
minimizing the reliance on extensive expert knowledge.  
Additionally, we parameterize translation results to generate reusable
translation rules to expedite future translations.

In summary, this paper makes the following contributions:
\begin{itemize}[noitemsep,topsep=0pt,leftmargin=3mm] 

\item We advocate for increased research in binary translation to support
software migration for new architectures like AArch64.
To this end, we introduce a novel static
assembly-to-assembly translation paradigm that utilizes
advanced binary rewriting techniques to produce portable binaries
with performance comparable to native counterparts.

\item We present an automatic binary translation pipeline that leverages LLMs to
construct the translation engine, drastically reducing the reliance on costly
expert knowledge and manual effort. Meanwhile, our approach simplifies
the translation task for LLMs by splitting code into snippets, allowing
for straightforward equivalence checking to ensure correctness and scalability.

\item We present an academic binary translation framework with performance that
rivals leading industrial tools. Evaluation shows that our research prototype,
\tool, significantly outperforms existing open-source solutions and even
surpasses a leading industrial dynamic binary translator, ExaGear, in some
real-world scenarios.
The translated binaries achieve near-native performance while remaining
portable. 
We also analyze the remaining runtime overheads, discuss \tool's
limitations, and suggest potential improvements for future research.

\end{itemize}

\section{Background}
\label{sec:preliminary}

\subsection{Binary Analysis and Reverse Engineering}
\label{subsec:re}

\noindent \textbf{Disassembly.}~
Disassembly is the foundational process of converting binary into human-readable
assembly. Two primary disassembly techniques are linear sweeping and recursive
traversal. Linear sweeping~\cite{de2000static} decodes all bytes in the
executable region sequentially,
while recursive traversal~\cite{kinder2010static} begins at a
designated entry point and decodes following control transfer instructions.
Recently, machine learning techniques have emerged, utilizing
neural networks to map binary bytes to assembly
instructions~\cite{yu2022deepdi,pei2020xda,ye2023d}. 
Despite decades of research, perfect disassembly remains elusive.
Disassemblers struggle with indirect jumps and calls, as they cannot statically
determine all control flow targets, often resulting in incomplete and inaccurate
outputs 
\cite{andriesse2016depth,kim2023reassembly}.

\begin{figure}[ht]
    \centering
    \includegraphics[width=0.78\linewidth]{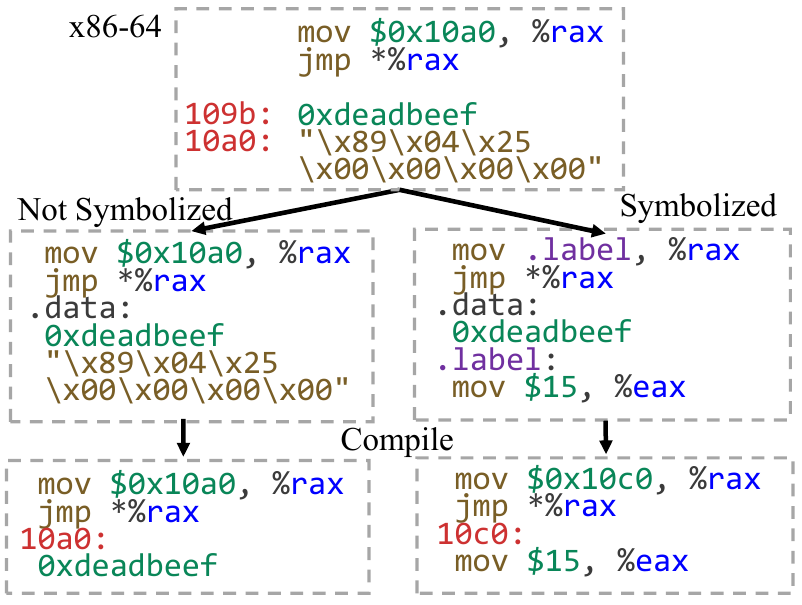}
    \caption{An example of symbolization.}
    \label{fig:symbolize}
\end{figure}

\noindent \textbf{Symbolization.}~
A key challenge in reverse engineering is the loss of symbolic information
during compilation. Programmer defined symbols such as functions, variables, and
labels are discarded, leaving only constant offsets in the binary.
As depicted in \F~\ref{fig:symbolize}, these symbols become indistinguishable
from other constants, leading to what is known as the \textit{symbolization
problem}~\cite{wang2015reassembleable,wang2017ramblr}. This poses a significant
hurdle for recompilation, as new addresses assigned by compilers may lead to
incorrect symbol references.

To address this, researchers have developed advanced \textit{reassemblable
disassembling}
techniques~\cite{wang2015reassembleable,wang2017ramblr,flores2020datalog,dinesh2020retrowrite,williams2020egalito}.
Notable approaches leverage binary analysis and relocation information in
position-independent code (PIC) to identify code and data
pointers~\cite{dinesh2020retrowrite,williams2020egalito}. They
focus on recovering lost symbols in disassembled code and have proven highly
reliable.
Recent advancements have even introduced sound binary
rewriting techniques~\cite{kim2025towards}, with many resulting tools rigorously
tested and validated~\cite{kim2023reassembly} in numerous real-world applications. 
\F~\ref{fig:symbolize} illustrates how proper symbolization enables
correct symbol references during recompilation.

\noindent \textbf{Symbolic Execution.}~Symbolic execution is a powerful program
analysis technique that verifies program properties for all possible
inputs~\cite{cadar2008klee,baldoni2018survey,poeplau2020symbolic}. Instead of
using concrete data, it executes a program with symbolic values and tracks them
throughout the program. When a branch is encountered, it explores both paths
separately, generating a set of symbolic constraints that represent the
conditions required to reach specific program states.   
These constraints can be evaluated by SMT
solvers~\cite{de2008z3,li2014symbolic} to check for
satisfiability and verify program properties. 
Symbolic execution has proven effective for analyzing
binary code~\cite{qi2024symfit,poeplau2021symqemu,wang2017angr}, 
offering a rigorous approach to verify translated binary code.

\subsection{Large Language Models}
\label{subsec:llm}

Recent breakthroughs in LLMs have demonstrated their effectiveness in various
natural language processing tasks like machine translation and text
summarization~\cite{chang2024survey,webson2021prompt,zhao2023survey}. Their
application has also extended to code-related tasks, such as code completion and
compilation~\cite{ziegler2022productivity,cummins2024meta,guo2024deepseek}. The
power of LLMs stems from their extensive training on high-quality data, which
cultivates sophisticated reasoning and inference capabilities for tasks that
traditionally required significant human expertise. 
Inspired by this, our work explores the potential of LLMs for
cross-architecture instruction translations, yielding promising results.
However, LLMs are known for their unpredictability; they can produce
plausible but incorrect outputs with no guarantee of
correctness~\cite{zhang2023siren,huang2023survey}. To mitigate this risk and
prevent the generation of erroneous binaries, we introduce a crucial rigorous
semantic equivalence check step, detailed in \S~\ref{subsec:design:se}.

\subsection{Binary Translation}
\label{subsec:bt}

Binary translation has been a long-standing topic of
academic research~\cite{song2019unleashing,wang2016general}. This paper focuses
specifically on user-level binary translation, which handles general
applications~\cite{bartholomew2006qemu,jiang2024system,wang2011coremu,d2017hypermambo,spink2020retargetable}, as opposed to system-level binary translation that targets
operating systems (OSes)~\cite{jiang2024system}.

\noindent \textbf{General Workflow.}~
A typical binary translator begins by disassembling the binary
software to extract its assembly code.
Then, depending on the translator's design,    
it can directly translate the assembly code to the host ISA, or lift assembly to a
platform-independent IR, which is then lowered to host assembly using
established compiler infrastructures (e.g., LLVM).
For efficiency, instructions are often organized and translated in translation
blocks (TBs)~\cite{bellard2005qemu,li2024tiaozhuan} that are typically stored in
a cache to avoid redundant translations.

\noindent \textbf{Terminology.}~In this paper, the term
\textit{binary translators} refers to tools for translating and
migrating binary software across architectures.
However, existing tools use distinct technical approaches with different
IRs and translation styles employed, 
significantly impacting their capabilities and performance.
To differentiate these
approaches clearly, 
we extend the terminology
developed in recent studies~\cite{liu2022sok,parzefall2024you} and categorize
them into four types:

\sparh{Concise IR Lifter.}~These tools aim to recover LLVM IR that closely
resembles the IR code generated by compiling source code. They \textit{lift}
low-level binary code to high-level, concise, and human-readable IR, attempting
to restore source-level information (e.g., variables, types, and symbols).
Notably, works like Mctoll~\cite{yadavalli2019raising}, RetDec~\cite{retdec},
and Rev.ng~\cite{di2017rev,di2018rev} have dedicated significant effort in this
direction. One subsequent advancement, Lasagne~\cite{rocha2022lasagne}, has
further enhanced Mctoll's capabilities for concurrent programs. 
Ideally, such lifters offer high flexibility, as the lifted IR can naturally
provide cross-architecture portability and can be readily analyzed and optimized
using established LLVM-based tools. 

However,
perfect IR lifting faces multiple
fundamental challenges, including incomplete control flow graph (CFG)
recovery~\cite{di2017rev,meng2016binary,shoshitaishvili2016sok} and the loss of
high-level abstractions like functions and variables in binary
code~\cite{he2018debin,xie2024resym}. 
Consequently, these tools often do not prioritize functionality preservation;
instead, they are
primarily used for automated program analysis, binary
similarity comparison~\cite{marcelli2022machine}, and reverse engineering tasks
like decompilation~\cite{mantovani2022convergence}.
Our evaluation confirms that they are not suitable for reliable
cross-architecture binary translation.

\sparh{Emulational IR Lifter.}~In light of the difficulties in recovering
high-level abstractions from binary code, another line of research focuses on
emulating the binary's semantics at the IR level. These lifters translate
each machine instruction into a sequence of IR statements that precisely
mimic its effects on a virtual CPU in lifted IR
code~\cite{liu2022sok,parzefall2024you}. For example,
McSema~\cite{dinaburg2014mcsema,mcsema} maintains a
dedicated structure in lifted IR to represent the CPU state, including registers
and flags.    
A similar paradigm is employed by dynamic lifters like
Instrew~\cite{engelke2020instrew,engelke2021efficient} and
BinRec~\cite{altinay2020binrec}. Two subsequent variants of BinRec, named
WYTIWYG~\cite{parzefall2024you} and Polynima~\cite{deshpande2024polynima}, 
while enhancing its support for stack-layout recovery and concurrent
programs, still fall into this emulational category.

Despite guaranteeing a high degree of functionality, this approach sacrifices the
benefits of LLVM IR as a strongly-typed IR. 
The lack of high-level semantic information, such as types, hinders automated
analysis and optimization~\cite{liu2022sok,parzefall2024you}. Furthermore, this
emulational strategy often introduces extra code in translated binary, resulting
in inefficiency~\cite{liu2022sok,parzefall2024you}. Given that static lifting
struggles with complete CFG recovery (McSema relies on the de facto
disassembler IDA Pro~\cite{ida} for CFG recovery and disassembly),
dynamic lifters like Instrew defer IR compilation and optimization to runtime,  
which further exacerbates performance issues.

\begin{table}[!tp]
	\centering
	\caption{Relevant binary translators investigated.} 
	\resizebox{0.99\linewidth}{!}{
		\begin{tabular}{ l | c | c | c | c | c }
			\hline 
            & \multirow{2}{*}{Type}   & IR &  Functional  & \multirow{2}{*}{Flexibility$^*$} & Primary \\
            &  & Used &  Fidelity  &  & Goal \\
            \hline
			Mctoll~\cite{yadavalli2019raising}  & Static  & LLVM  &  Low & High & \multirow{3}{*}{\shortstack{Readable\\ Concise\\ IR}}  \\
            \cline{1-5}
            RetDec~\cite{retdec}  & Static  & LLVM  &  Low & High & \\
            \cline{1-5}
            Rev.ng~\cite{di2017rev}  & Static  & LLVM  &  Low & High & \\ 
			\hline
            \hline
            McSema~\cite{mcsema}  & Static  & LLVM  &  Medium & Medium & \multirow{3}{*}{\shortstack{High-\\ Fidelity\\ IR}}  \\
            \cline{1-5}
            Instrew~\cite{engelke2020instrew}  & Dynamic  & LLVM  &  High & Medium & \\
            \cline{1-5}
            BinRec~\cite{altinay2020binrec}  & Dynamic  & LLVM  &  High & Medium & \\
			\hline
            \hline
			\multirow{2}{*}{QEMU~\cite{bartholomew2006qemu}}  & \multirow{2}{*}{Dynamic}  & \multirow{2}{*}{TCG}  &  \multirow{2}{*}{High} & \multirow{2}{*}{Low} & Reliable \& \\ 
			  &   &   &   &  & Portable \\ 
            \hline
            \hline
			ExaGear~\cite{exagear}  & Dynamic  & None  &  High & Low & \multirow{2}{*}{\shortstack{High\\ Performance}} \\ 
			\cline{1-5}
            \textbf{\tool}  & \textbf{Static}  & None  &  High & Low &  \\ 

            \hline
		\end{tabular}
	}
	\label{tab:related}
\begin{flushleft}
\noindent \footnotesize{$^*$ In this context, flexibility is defined as
the ease of porting to new architectures and applying standard compiler analyses
and optimizations.}\\
\end{flushleft}
\end{table}

\sparh{Low-level IR Translator.}~
This category of
works~\cite{bartholomew2006qemu,quynh2015unicorn}, which includes CPU emulators
like QEMU, forgoes the cross-architecture
benefits of LLVM IR in favor of a lower-level IR that more closely mirrors
machine code.
For decades, QEMU~\cite{bartholomew2006qemu} has been widely adopted in
industry and academia due to its high reliability.

Recent studies, however, indicate that QEMU can be over 10$\times$
slower than native execution~\cite{xie2024instruction}. 
This performance penalty stems from a fundamental trade-off between scalability
and portability. 
To support a variety of architectures, QEMU uses a lower-level IR (TCG
IR)~\cite{bellard2005qemu,bartholomew2006qemu}. Each CPU instruction is
decomposed into several micro-operations in TCG IR, which are implemented in
high-level programming languages and compiled into native host machine code. 
While this shields developers from the intricacies of complex ISAs and
hardware specifics, it limits low-level
optimization opportunities and introduces inefficiencies compared to
hand-tuned assembly, leading to substantial overhead.

\sparh{Assembly Translator.}~To achieve maximum performance, industrial
dynamic translators like ExaGear~\cite{exagear} bypass IRs and
translate guest instructions directly into native host instructions, leveraging
hardware features and delivering near-native performance. 
The downside is that developing the translation logic is labor-intensive,
error-prone, and typically tailored for specific platforms, limiting portability
and reusability. Moreover, the dynamic paradigm they employed also introduces
substantial runtime overhead.

\tool\ mitigates these issues by automating the translation process using LLMs. 
Its static paradigm allows us to incorporate LLMs and rigorous verification into
the translation process.
Moreover, by decoupling the translation framework from
translation rules, we transform translation into a data-driven process. 
As a result, developing the translator requires minimal expert knowledge, as
rules are mostly extracted from LLM-generated examples. 

It is important to note, however, \tool is not intended to replace the
mature binary translators developed by industry over the years. Binary
translation remains a highly complex field requiring substantial engineering
effort (e.g., QEMU consists of over one million LoC, and Huawei has spent more
than five years refining ExaGear). Because there are still many unresolved
problems requiring further research, this work primarily contributes new
concepts and a novel translation paradigm to the community.

\smallskip
\noindent \textbf{Previous Works to Compare.}~
To position our work, 
\T~\ref{tab:related} presents a qualitative synthesis and taxonomy of
representative translators.
Specifically, the classification is grounded from a recent
research~\cite{liu2022sok}, whose observations have been confirmed by subsequent
studies~\cite{parzefall2024you,zhou2024plankton}. We adopt this established
terminology and taxonomy while expanding our scope to include four additional
tools (Rev.ng, Instrew, QEMU, and ExaGear) for a more thorough discussion.
Within this taxonomy, functional fidelity refers to the strictness of
semantic preservation. High fidelity ensures full recompilability, while low
fidelity prioritizes static analysis and code comprehension over executability.
Although McSema faithfully lifts all individual instructions for small-scale
benchmarks~\cite{dasgupta2020scalable}, its performance in our evaluation
justifies a medium fidelity rating (as detailed in \T~\ref{tab:rk3588-1}). 
Flexibility follows a similar categorization. High flexibility indicates the
recovery of strongly-typed IR compatible with standard compiler passes, while
low flexibility refers to methods tied to hardware specifics such as direct
assembly translation. Emulational IR lifters occupy a middle ground, producing
IR that allows for cross-target lowering but lacks support for standard compiler
analysis.

\T~\ref{tab:related} highlights the functional and structural limitations of
current approaches. Our investigation found that while BinRec shows promising
performance, it is designed for same-architecture recompilation and optimization
and is unsuitable for cross-architecture translation because it must concatenate
the lifted binary with the original.
Concise IR lifters, despite backing from notable technology companies, often
fail to deliver fully functional results (more details of their
failures in the Appendix~\ref{subsec:appendix:lifter}). In fact, tools like
RetDec and Rev.ng primarily aim at software decompilation, treating translation
failures and semantic mismatches as low-priority issues~\cite{revng-issue}. 
Consequently, our evaluation mainly compares \tool with state-of-the-art in each
category: McSema, Instrew,
QEMU, and ExaGear. 
\S~\ref{sec:motivation} further discusses more design choices and 
rationales of \tool, aiming for high reliability and runtime efficiency. 

We also acknowledge recent advancements in this domain that solve specific
complexities, such as concurrent
programs~\cite{rocha2022lasagne,deshpande2024polynima,gouicem2022risotto} and
stack-layout recovery~\cite{parzefall2024you}, in modern software. They are not
separately listed as they are primarily built upon works already covered in
\T~\ref{tab:related}, with their core designs remaining unchanged. We further
discuss them in \S~\ref{sec:related}.

\section{Motivation}
\label{sec:motivation}

As discussed in \S~\ref{subsec:re}, static disassembly struggles to resolve
indirect control flow targets and symbols missed during compilation.
While some static translators employ mature commercial disassemblers
like IDA Pro~\cite{ida} to partially mitigate these issues~\cite{mcsema}, they
remain brittle in general scenarios. 
This unreliability discourages the adoption
of static disassembly in translators where execution correctness is paramount. 
Consequently, industrial binary translators typically rely on dynamic disassembly,
which necessitates a dynamic translation paradigm and introduces significant
runtime overhead, 
preventing them from producing portable binaries, i.e., a guest binary can only
execute on platforms with the translator installed.

To overcome these fundamental limitations, we introduce a new
translation paradigm centered around the following major conceptual
contributions:

\noindent \sparh{1) Static Assembly-to-Assembly Translation.}~As noted
in \S~\ref{subsec:re}, recent advancements in binary rewriting present a
\textit{unique opportunity} for static binary translation. Leveraging these
binary rewriting techniques, we propose a novel static translation paradigm that
translates assembly code, 
bypassing the challenges of IR lifting and ensuring
high reliability.
Furthermore, the static paradigm reduces runtime overhead and produces portable,
native-like binaries, opening new avenues for research and downstream
optimization.  

\begin{figure*}[htp!]
  
  \centering
  \includegraphics[width=0.80\linewidth]{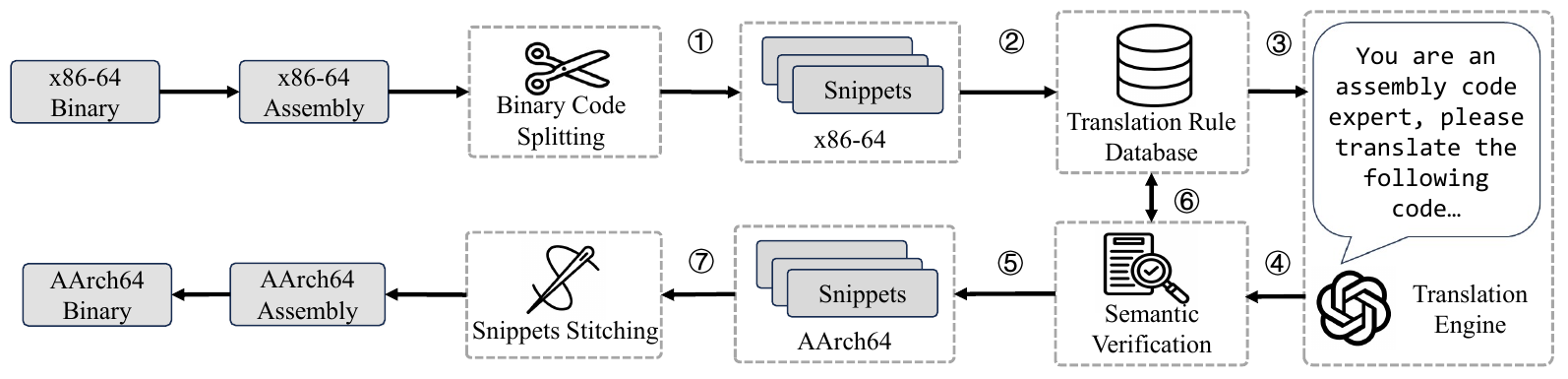}
  
  \caption{Overview of \tool.}
  
  \label{fig:overview}
\end{figure*}

\noindent \sparh{2) Automated Translator Development.}~State-of-the-art industrial
binary translators like Rosetta 2~\cite{apple-rosetta} and
ExaGear~\cite{exagear} are closed-source and tailored to specific hardware and
OSes, lacking support for general AArch64 processors. Meanwhile,
existing open-source
translators~\cite{mcsema,engelke2020instrew,altinay2020binrec} are unsuitable
for our static assembly-to-assembly paradigm because they lift assembly to IRs
(as introduced in \S~\ref{sec:preliminary}), which inevitably introduces
unnecessary overhead.
Building a new translator from scratch is a formidable task that requires
a tremendous engineering effort. The absence of open, efficient translators for
general AArch64 platforms hinders further research. 

To establish a practical foundation for future research, we introduce a novel
binary translation pipeline. 
Specifically, we propose using LLMs to automatically generate translation
results, replacing the need for human experts.
This approach is economically viable due to the affordability of modern LLMs.
Furthermore, since many instruction variants share similar translation
patterns, previously generated translations can be parameterized and reused to
accelerate future translation.

\noindent \sparh{3) Scalable Semantic Verification.}~CPU instruction semantics are
often intricate, and even experienced developers may mistranslate
subtle semantic details. 
Modern ISAs, particularly complex instruction set computing (CISC) architectures
like x86-64, contain a vast number of instructions, forcing developers to
spend significant time consulting manuals and designing test cases to ensure
accurate translation. 
This is a tedious and error-prone process that still does not guarantee correctness.
While methods like symbolic execution can rigorously verify binary code, they
often face scalability issues due to path explosion~\cite{schwartz2010all}.

Verifying translation correctness 
is a difficult but essential challenge. 
To address this, we introduce an efficient static verification method. 
Instead of verifying an entire complex program, we split assembly code into
small, manageable snippets that can be easily verified with symbolic execution
and then reassembled. This method provides instant feedback
to human experts or LLMs, 
streamlining the development of robust binary translators.

\section{Design}
\label{sec:design}

This work formulates a new translation paradigm, outlining a concrete methodology
for developing a high-performance, portable binary translator empowered by LLMs.
We present \tool, a static binary translator for migrating x86-64 software to
AArch64 platforms. This section presents a design overview and then examines
each component in detail. 

\subsection{Overview}
\label{subsec:design:overview}

As shown in \F~\ref{fig:overview}, the translation process starts with standard
disassembly. 
What distinguished our approach is the use of advanced binary rewriting
techniques to transform disassembled code into a symbolic, reassembleable form. 
We then split the binary code into snippets (\circled{1}), with each snippet
representing a code block in a Static Single Assignment (SSA)-like
format~\cite{cytron1989efficient,cytron1991efficiently,van2007static}. 
For clarity, we use ``SSA'' throughtout the paper to refer to our adapted
concept, which differs slightly from the traditional SSA used in compiler
research (as explained in \S~\ref{subsec:design:split}). The SSA representation
enables independent translation of each snippet. 
For each snippet, we first query the translation rule database for existing,
reusable translations (\circled{2}). If no cached translation can be
reused, our LLM-powered engine translates the snippet
(\circled{3}); otherwise, the existing translation
rule in the database
is applied. Once snippets are translated, we employ symbolic verification to
ensure the semantic correctness of the translated snippets (\circled{4}).
Finally, we stitch translated snippets into complete, compilable assembly code
(\circled{5} and \circled{7}). Moreover, to minimize the cost of repeatedly
querying the LLM for similar code, we further parameterize translation results
into reusable rules for future translations (\circled{6}). The following
subsections elaborate on each of these components.

\subsection{Binary Code Splitting}
\label{subsec:design:split}

Translating entire programs or functions poses significant hurdles to LLMs and
verification. 
The extensive context length required, coupled with intricate data and
control flows in assembly code, often exceeds the reasoning capabilities of LLMs.
Additionally, verifying the
correctness of large-scale translated code is a difficult task. 
Drawing inspiration from the divide-and-conquer strategy~\cite{smith1985design},
we propose a fine-grained translation approach. Our method focuses on
translating small code snippets formatted in SSA form. The snippets are
generally concise enough for LLMs to comprehend and process. The SSA form
simplifies data and control flows, minimizing potential errors 
introduced by LLMs and enabling the semantic verification of the translation.

\parh{Adapted SSA.}~ Through static analysis, we break down each basic
block into SSA snippets. Here, SSA requires that within each snippet, a variable
is assigned a value at most once (i.e., single assignment). Unlike the traditional
SSA concept used in compiler research, our adaptation includes cases where a
variable might be used before definition. 
This distinction arises because, unlike high-level languages, variables in
assembly (such as registers and memory) are inherently defined by the hardware
state and are never truly undefined. 

\begin{algorithm}[h]
\footnotesize
\caption{Splitting Basic Block to SSA Snippets}
\label{alg:ssa-split}

\SetKwProg{Fn}{Function}{}{}
\SetKwFunction{FSplit}{Split}
\SetKwFunction{FGetDef}{GetDefinedVariables}

\Fn{\FSplit{$BB_{instrs}$: a list of instructions from a basic block}}{
    $S \gets \emptyset$\;
    $CurrentSnippet \gets []$\;
    $AssignedVars \gets \emptyset$ \tcp*{variables assigned in $CurrentSnippet$}
    
    \ForEach{$instr$ \textnormal{in} $BB_{instrs}$}{
        $DefinedVars \gets \FGetDef{instr}$\;
        
        \eIf{$AssignedVars \cap DefinedVars \neq \emptyset$}{
            Add $CurrentSnippet$ to $S$\;
            $CurrentSnippet \gets [instr]$\;
            $AssignedVars \gets DefinedVars$\;
        }{
            Append $instr$ to $CurrentSnippet$\;
            $AssignedVars \gets AssignedVars \cup DefinedVars$\;
        }
    }
    
    \If{$CurrentSnippet$ is not empty}{
        Add $CurrentSnippet$ to $S$\;
    }
    
    \Return $S$ \tcp*{Output: A list of SSA snippets}
}

\Fn{\FGetDef{$instr$}}{
    $D \gets \emptyset$\;
    \If{$instr$ writes to memory}{
        \tcp{For simplicity, all memory is treated as one variable.}
        $D \gets D \cup \{\text{MEMORY}\}$\;
    }
    $R_{dest} \gets \text{destination registers of } instr$\;
    $D \gets D \cup R_{dest}$\;
    \Return $D$\;
}
\end{algorithm}

\parh{Splitting.}~This step starts with symbolized assembly code. The
splitting process is guided by the core principle of single assignment to reduce
complexity. As illustrated in \A~\ref{alg:ssa-split}, we linearly scan
the assembly instructions (lines 6, 13), tracking which variables are assigned
at each instruction. Whenever a previously assigned variable is rewritten
(line 7), we split preceding instructions into a snippet and reset the track
(lines 8--10).
This process continues until the end of a basic block, where control flow
transfers naturally conclude a snippet. 
Due to the intangibility of statically determining the address of memory
accesses, we treat all memory as a single variable for simplicity. This means
each snippet can have at most one memory access. This step yields a
list of SSA snippets ready for independent translation.

\subsection{LLM-Empowered Translation}
\label{subsec:design:llm}

Each x86-64 snippet is then translated to AArch64 individually with an LLM.
However, directly translating assembly code with LLMs can be error-prone,
particularly with complex CISC ISAs like x86-64. While LLMs generally possess
knowledge of relevant instructions, they can produce incorrect translations 
due to insufficient context or limited reasoning capabilities.
To meet the strict correctness requirements of binary translation, we design a
domain-specific prompt architecture that adapts recent advances in prompt
engineering to the constraints of assembly language.

\begin{figure}[htb]
  \centering
  \includegraphics[width=0.95\linewidth]{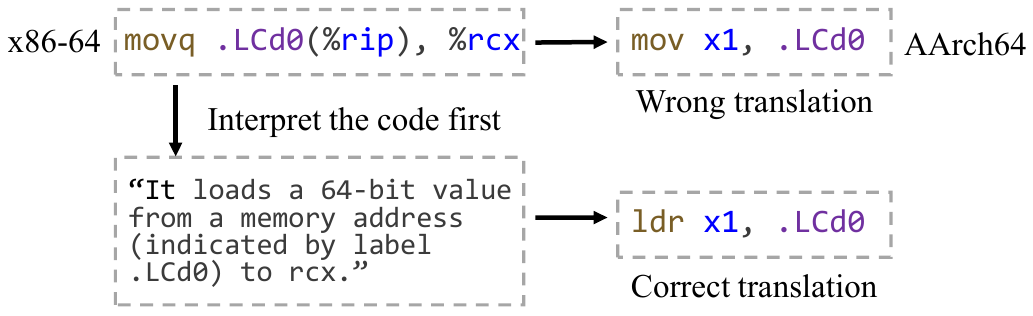}
  
  \caption{An example of semantic-grounded reasoning.}
    
  \label{fig:example-design-1}
\end{figure}

\noindent \textbf{Semantic-Grounded Reasoning.}~First, LLMs can exhibit
shallow reasoning, opting for superficial interpretations rather than performing
a deep semantic analysis of the code. 
To address this issue,
we implement a specialized task-splitting strategy, effectively adapting
chain-of-thought (CoT) prompting~\cite{wei2022chain} to the binary domain, which
instructs the LLM to interpret the assembly code before translating it. This
preliminary interpretation step grounds the LLM's understanding of code's
semantics, promoting more accurate and deliberate reasoning. We strategically
choose not to verify this intermediate interpretation, acknowledging that
hallucinations can occur at any stage. Instead, we treat interpretation as an
auxiliary cognitive scaffold to prime the model, focusing our rigorous formal
verification efforts exclusively on the final executable output (see
\S~\ref{subsec:design:se}). For example, as shown in
\F~\ref{fig:example-design-1}, the LLM initially misinterprets an instruction as
moving
the address of label \texttt{.LCd0} into \texttt{x1}. However, by first
prompting the LLM 
for a semantic interpretation,
the LLM can produce the correct translation, which loads a 64-bit integer from
the memory indicated by \texttt{.LCd0}.

\begin{figure}[h]
  \centering
  \includegraphics[width=0.97\linewidth]{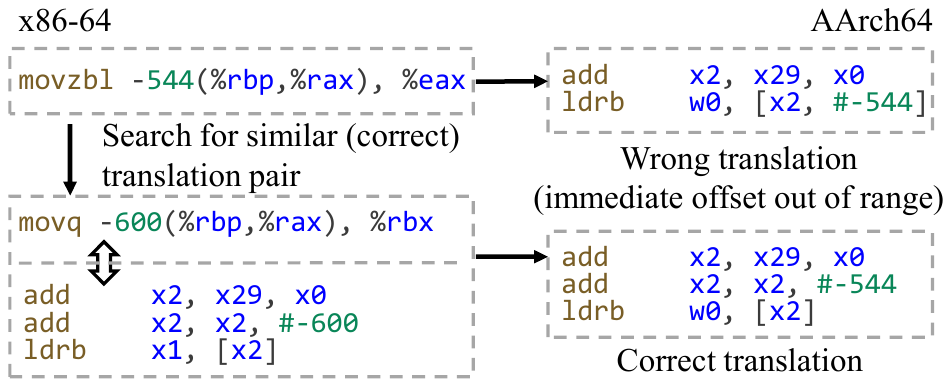}
  
  \caption{An example of RAG-enhanced translation.}
    
  \label{fig:example-design-2}
\end{figure}

\noindent \textbf{RAG-Enhanced Translation.}~Second, LLMs may generate
syntactically invalid code despite correct semantics understanding,
often due to insufficient context. 
We repurpose retrieval-augmented generation
(RAG)~\cite{cuconasu2024power,gao2023retrieval} to solve the brittleness issue
of assembly code synthesis.
We retrieve similar, correct translation examples from our database based on
text similarity and inject them as architectural context. 
These examples provide the LLM with precise, compilable syntax patterns, helping
LLM to produce valid code.
\F~\ref{fig:example-design-2} shows an example where the initial translation
(top right) is non-compilable because an immediate value (\texttt{-544}) exceeds the allowable
range. By providing a textually similar, valid example, the LLM  learns
the correct pattern and generates a valid translation (bottom right).

\noindent \textbf{Iterative Refinement.}~Finally, 
to fix any remaining errors,
we introduce an iterative
process to refine and correct the results.
A key contribution of our approach is replacing standard heuristic-based 
refinement with a feedback loop grounded in formal verification.
Specifically, we 
perform a rigorous semantic verification using the \texttt{angr}
symbolic execution engine~\cite{wang2017angr} and the Z3 solver~\cite{de2008z3}
(details in \S~\ref{subsec:design:se}). If any errors or semantic
inconsistencies are found, the precise feedback (i.e., formatted output
from the symbolic execution engine) is passed back to the LLM.
The LLM is then prompted to refine the translation based on this feedback,
effectively combining the reasoning ability of LLMs with the rigor of
formal proofs. The newly refined translation is then verified with symbolic
execution again. This iterative cycle continues until the translation passes
semantic verification or a predefined iteration limit (set to 5) is reached. 

\begin{figure}[t]
  \centering
  \includegraphics[width=0.78\linewidth]{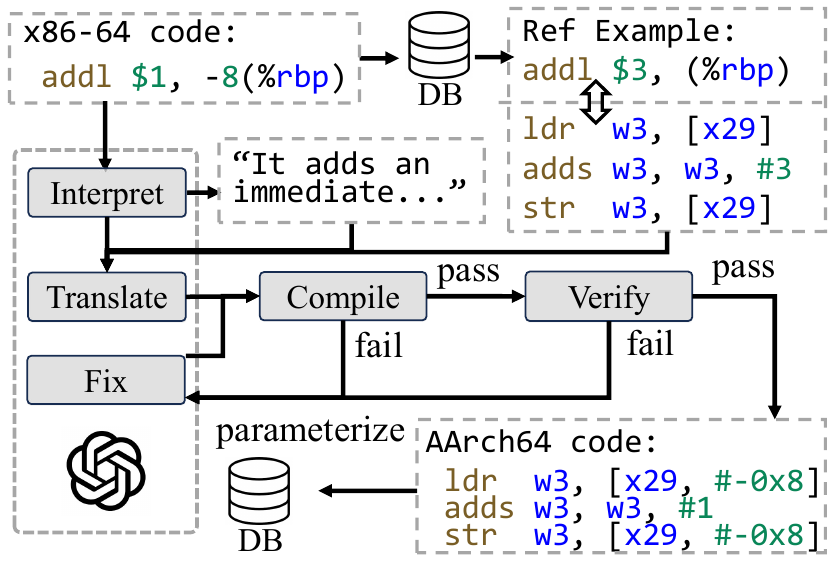}
  
  \caption{Workflow of iterative refinement.}
    
  \label{fig:translate}
\end{figure}

As a result, we establish an iterative pipeline for our
translation engine, as shown in \F~\ref{fig:translate}. While the
interpret, translate, and fix steps are LLM-powered,
we use the
GCC compiler and 
\texttt{angr} for compilation and verification. Our evaluation confirms that 
this specialized integration of prompt engineering and formal
verification 
forms a necessary innovation to enable reliable, LLM-based binary translation.

\subsection{Semantic Verification}
\label{subsec:design:se}

The semantic verification process ensures
that the translated code is semantically equivalent to the original. Two
snippets are considered semantically equivalent if they 
always produce identical outputs for the same inputs. 

\noindent \textbf{Variables.}~
This equivalence requires a one-to-one mapping between input and output
variables, where corresponding variables hold the same value after execution.
Variables in assembly code include
registers and memory, while memory is treated as a special variable that
encompasses both its value and the address being accessed.
Furthermore, instruction execution often updates the flags register,
which subsequent instructions (especially branches) may utilize. Since
\texttt{angr} treats flags as specialized registers, we can verify them using
the same methodology for general-purpose registers.
While x86-64 defines six status flags, two of them (AF and PF) are primarily
retained for backward compatibility~\cite{guide2011intel,intel8086}. Our
analysis confirmed that these two flags are unused in our evaluation set. The
remaining four flags have direct counterparts in AArch64 and can be mapped
reliably.

\begin{figure}[h]
  \centering
  \includegraphics[width=0.87\linewidth]{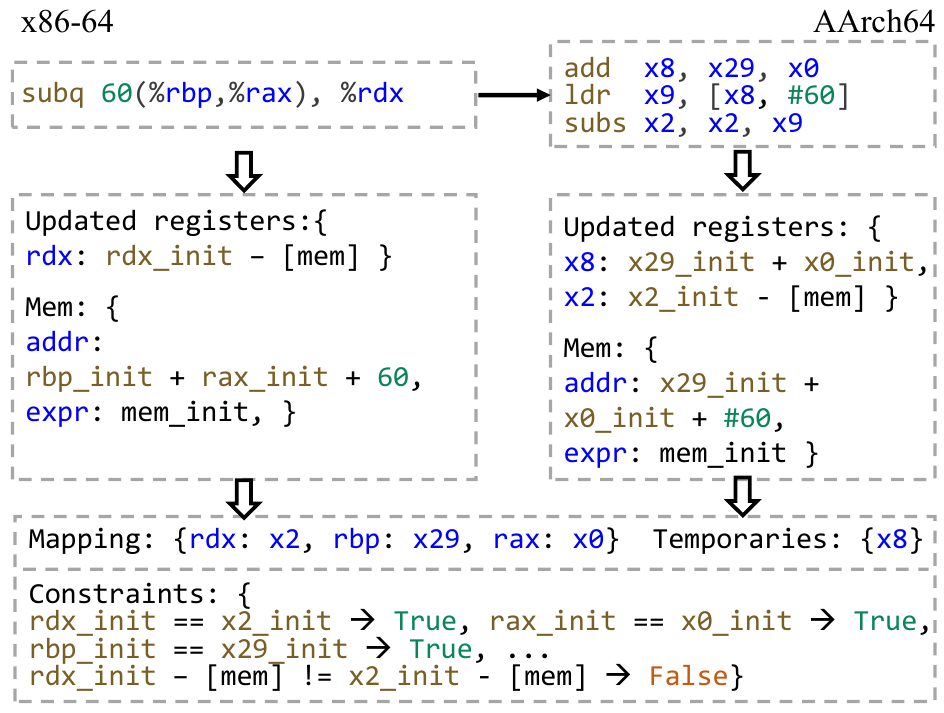}
  
  \caption{An example of symbolic execution results.
  }
  
  \label{fig:example-design-se}
\end{figure}

\noindent \textbf{Symbolic Execution.}~To verify semantic equivalence,
we first perform symbolic execution on both the x86-64 and translated snippets.
We assign initial symbols to all input variables and conduct symbolic execution
to derive symbolic expressions for each output. Crucially, the snippets
are ensured to have only one valid execution path without branches. This design
choice deliberately bypasses the path explosion problem that complicates
symbolic execution in general scenarios. \F~\ref{fig:example-design-se}
demonstrates an example of symbolic execution results, where initial symbols are
marked with an \texttt{\_init} suffix.
The verification requires finding a valid mapping
between registers that satisfies the aforementioned equivalence property. 
Using brute-force search, we explore possible register mappings and employ the
Z3 constraint solver~\cite{de2008z3} to check if a given mapping satisfies the
equivalence property. The small number of variables makes this search
computationally efficient. If a valid mapping is found, the two snippets are
deemed semantically equivalent.

\noindent \textbf{Equivalence Checking.}~Assuming the x86-64 snippet
contains variables $V_{A} = \{ v_{A_1}, v_{A_2}, v_{A_3}, \dots, v_{A_{nA}} \}$,
and the AArch64 snippet uses variables $V_{B} = \{ v_{B_1}, v_{B_2}, v_{B_3},
\dots, v_{B_{nB}} \}$, let $E(v_{i})$ denote the symbolic expression of variable
$v_{i}$, with $E_{init}(v_i)$ as its initial symbolic expression. Given a
mapping $M: V_{A} \rightarrow V_{B} $, for each variable $v_{i} \in V_{A}$, we
first establish the initial conditions by adding the constraint $E_{init}(v_{i})
= E_{init}(M(v_{i}))$ to the solver. This ensures that corresponding
variables start with the same initial symbolic expression.  
After that, for each variable $v_{i} \in V_{A}$, we check if $E(v_{i}) \neq
E(M(v_{i}))$ is unsatisfiable with the solver. If it is, we add
$E(v_{i}) = E(M(v_{i}))$ to the solver and proceed to the next
variable. This ensures that symbolic expressions of corresponding
variables \textit{must} be equal after execution. Once all
variables are validated, the mapping is considered to satisfy the equivalence
property. 
This process naturally handles temporary registers that may be introduced in the
AArch64 snippet, as x86-64 instructions are often decomposed into multiple
simpler RISC operations.

\subsection{Translation Rule Parameterization}
\label{subsec:design:rule}

\begin{figure}[t]
  \centering
  \includegraphics[width=0.95\linewidth]{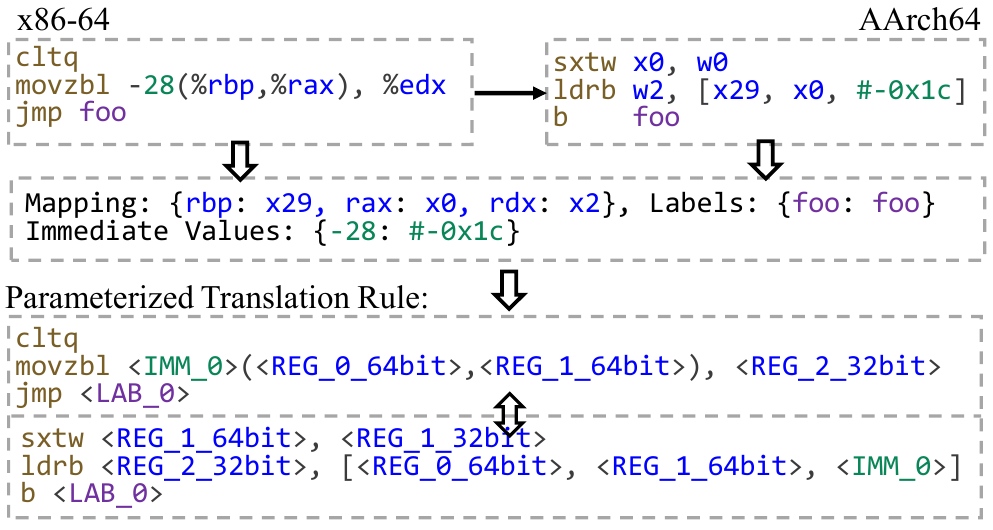}
  
  \caption{An example of a parameterized translation rule.}
  
  \label{fig:example-rule}
\end{figure}

Querying an LLM is time-consuming and inefficient, especially
for repeated similar code. To address this, we parameterize verified
translation results into reusable rules.

\noindent \textbf{Parameterization.}~
Specifically, we observed that snippets with similar structure, albeit
with different operands, 
often yield translation results with the same sequence of AArch64 instructions.
Therefore, we abstract concrete operands (e.g., registers and immediate values)
into parameters, extracting generalized snippet skeletons. This allows the reuse
of a verified translation rule for multiple similar x86-64 snippets without
repeated LLM queries.
Furthermore, the parameterized translation rules are context-independent and can
guarantee consistent functionality when reassembled. 

\F~\ref{fig:example-rule} illustrates an example of a parameterized translation
rule. The translation at the top is semantically correct and verified.
We replace its concrete operands with unique parameter identifiers (composed of
its type and ID) based on the mapping derived from verification. 
These parameterized rules, along with
original snippets, symbolic verification results, and mappings, are stored in
the rule database. 

\noindent \textbf{Database Querying.}~Before translating a new x86-64
snippet, we first parameterize it and query the database for matched rules.
A match requires that the 
snippet's instruction, parameter types, and sequence are identical to a rule in
the database.
Upon finding a match, the rule is applied by replacing parameters with concrete
operands. Otherwise, the snippet is sent to the LLM. 
For consistency, we conduct the same
parameterization
process on translation results produced by
LLMs.
In practice, as the database grows, 
these rules increasingly supplant the
role of LLMs, shifting translation into an efficient data-driven
process. Consequently, \circled{3} and \circled{4} in \F~\ref{fig:overview}
become less frequent, with most snippet translations being handled by reusing
rules in the database (\circled{6}),
reducing the time and cost of LLM queries.

\noindent \textbf{Concretization.}~Once all snippets are translated,
the resulting rules, whether from the database or generated by parameterizing
LLM outputs, need to be concretized before they can be stitched into complete
assembly code. 
This process reinserts specific operands  
back into the translated snippets.
While immediate values and labels can be directly extracted from x86-64
snippets, register concretization demands a consistent mapping across snippets
(i.e., a specific x86-64 register must always map to the same AArch64 register).
For simplicity, we employ a fixed mapping to define the correspondence between
x86-64 and AArch64 registers, as detailed in 
\T~\ref{tab:reg-map}.
This is feasible without causing register conflicts because AArch64
offers more general-purpose registers than x86-64. The remaining AArch64
registers are utilized for temporary values introduced during translation.

\begin{table}[h]
	\centering
	\caption{The mapping used to concretize translation rules.}
	
  \resizebox{0.69\linewidth}{!}{

		\begin{tabular}{l | c | c }
			\hline 
			\textbf{x86-64} & \textbf{AArch64}  &   \textbf{description} \\ 
			\hline
			rdi & x0 & \multirow{6}{*}{Function arguments}\\
      rsi & x1 & \\
      rdx & x2 & \\
      rcx & x3 & \\
      r8 & x4 & \\
      r9 & x5 & \\
      \hline
      rax & x8 & Return value\\  
      \hline
      rbx  & x9 & Callee saved\\
      r10-r15  & x20-x25 & temporary registers\\
      \hline
      rbp & x29 & Frame pointer \\
      \hline
      rsp & sp & Stack pointer \\
      
      \hline
		\end{tabular}
	}
	\label{tab:reg-map}
  
\end{table}

\subsection{Snippets Stitching}
\label{subsec:design:connect}

The final step is to stitch concretized snippets into
an executable
assembly program. This offline process introduces no runtime overhead and
primarily involves reconstructing the program's control flow and managing
Application Binary Interface (ABI) differences 
with function calls.

\uparh{Control Flow Reconstruction.}~
We arrange the translated snippets in the original
sequence as their x86-64 counterparts. We iterate through 
each basic block and emit AArch64 assembly code for each snippet. Jumps
and branches 
are resolved by mapping all target
addresses to appropriate labels.
For instance, an \texttt{jne $.label$} is translated into a corresponding
\texttt{b.ne $.label$} in AArch64, where \texttt{$.label$} points to the
equivalent location in the new code. 

\begin{figure}[t]
  \centering
  \includegraphics[width=0.85\linewidth]{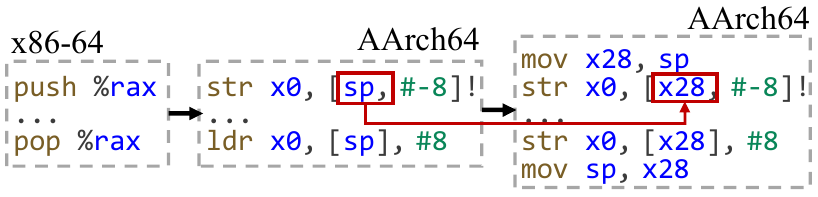}
  
  \caption{Stack Alignment Management.}
  \label{fig:example-stack-align}
  
\end{figure}

\uparh{Stack Alignment Management.}~Moreover, AArch64 enforces stricter stack
alignment requirements than x86-64. 
A common issue arises when the prologue code of an x86-64 function pushes an 8-byte
value, which can misalign the 16-byte boundary required by AArch64.
To resolve this, we employ a standard AArch64 assembly programming
technique~\cite{aarch64-push}. 
Upon entering a function, we 
substitute the stack pointer (\texttt{sp}) with a free register (e.g.,
\texttt{x28}). 
As shown in \F~\ref{fig:example-stack-align}, using \texttt{x28} in the store
(\texttt{str}) instruction allows the original stack layout to be preserved
while ensuring that \texttt{sp} remains 16-byte aligned. The stack pointer is
restored upon exiting the function, maintaining proper stack discipline.

\smallskip
\parh{Correctness.}~The correctness of the stitched program hinges on
the principle of \textit{compositional
verification}~\cite{cobleigh2003learning}. 
Each x86-64 snippet is translated into a semantically equivalent AArch64
snippet. The use of SSA form makes data flow explicit through def-use chains,
allowing each snippet to be treated as a verified, context-independent function.
Since each translated snippet produces identical output as the original
snippet, the correctness of stitched programs can be reasoned.
However, the correctness guarantee relies on two main assumptions:

\uparh{1) No Implicit Cross-Snippet Dependencies.}~We assume that there are no
hidden side effects or dependencies between snippets that are not captured by
the explicit control and data flow. This holds for standard application code but
excludes highly specialized cases like self-modifying code.

\uparh{2) Mediated ABI Compatibility.}~We assume that external libraries are
effectively managed, and ABI incompatibilities are resolved with additional
expertise. That requires all necessary libraries to be available on the AArch64
host, and our hooking mechanisms (introduced below) are applied to function calls
to mediate the
incompatibilities.

\begin{figure}[t]
  \centering
  \includegraphics[width=0.94\linewidth]{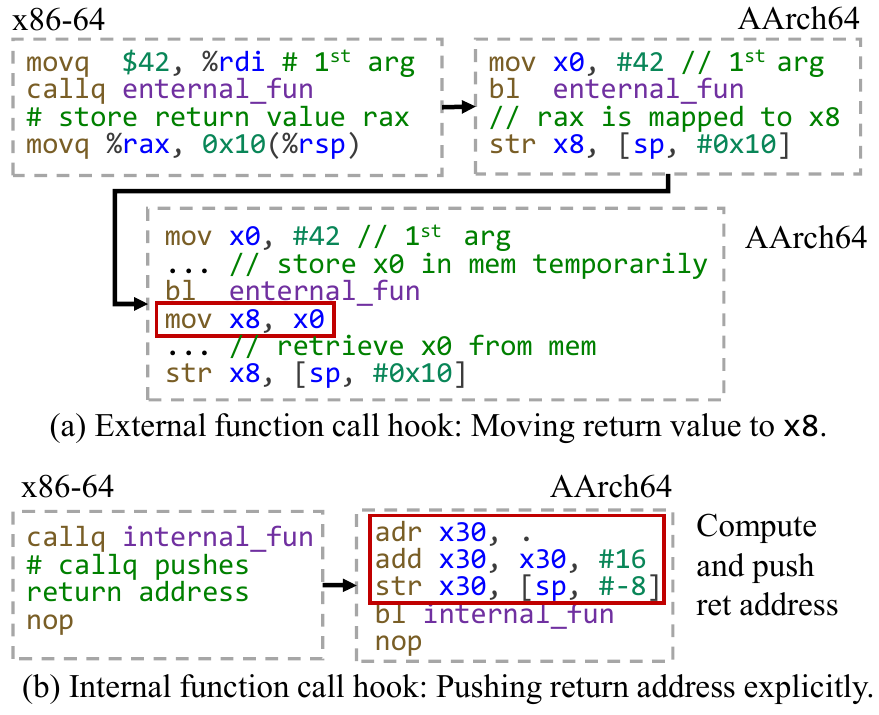}
  
  \caption{Examples of extra code inserted.}
  \label{fig:example-issues}
  
\end{figure}

\smallskip
\parh{Function Call Hooks.}~While stitching, we hook
calls to resolve ABI differences between the x86-64 and AArch64, which arise
from their distinct architectural designs.
While mitigating such issues requires
domain expertise and extra code,
this is 
a standard practice
for cross-architecture binary translation and
commonly seen in related binary analysis works~\cite{kim2025towards,yadavalli2019raising,meng2016binary,dinesh2020retrowrite}.
Specifically, our hooks address inconsistencies in calling conventions
for external calls and
return address handling for internal calls. 

\uparh{Calling Convention.}~Since translated code follows the x86-64
calling convention, it requires no special handling for arguments within
translated functions. 
However, differences in the ABI must be addressed at the boundaries between
translated and native code (e.g., external library calls).
First, the return value and first
argument in x86-64 are in separate registers (\texttt{rax} and
\texttt{rdi}), whereas in AArch64, both are handled by the \texttt{x0}
register. Our implementation maps \texttt{rax} and \texttt{rdi} to
different registers (\texttt{x8} and \texttt{x0}, respectively). 
As demonstrated in \F~\mysubref{fig:example-issues}{a}, this is addressed by
preserving and retrieving the first argument before and after a function call,
and then transferring the return value from \texttt{x0} to \texttt{x8} after the
call.
Second, x86-64 passes the first six arguments via registers
(\texttt{rdi}-\texttt{r9}), with subsequent arguments on the stack.
AArch64, however, uses registers \texttt{x0}-\texttt{x7} for the first eight
arguments. Our fixed register mapping already aligns x86-64's
\texttt{rdi}-\texttt{r9} with AArch64's \texttt{x0}-\texttt{x5}. To ensure
correct argument passing for functions with more than six arguments, we move the
seventh and eighth arguments from the stack into registers \texttt{x6} and
\texttt{x7} before the call.

\uparh{Return Address.}~
Additionally, x86-64 and AArch64 also manage
return addresses differently. 
In x86-64, the \texttt{call} instruction automatically 
pushes the return address onto the stack, while in
AArch64, the \texttt{bl} instruction stores the return address in the link
register (\texttt{x30}). 
Thus, an internal function call from translated code will leave the return
address in \texttt{x30}, which can be overwritten by subsequent calls.
Our solution, shown in \F~\mysubref{fig:example-issues}{b}, hooks the call to
explicitly push the return address 
to preserve it and pop it back into \texttt{x30} before returning, ensuring the
integrity of the call stack.

\smallskip
\parh{Threats to Validity.}~The issues resolved in the stitching process are
unique for assembly-level translators, as lifters typically compile lifted IR
into binary without diving deep into architectural differences. 
While domain expertise is required, the core problems and solutions are common
and well-documented~\cite{arm-abi,arm-call}. 
Nevertheless, the validity of stitching faces potential threats:

\uparh{Complex Data Types.}~While we address common ABI discrepancies, issues
may arise with specialized data types (e.g.,
floating-point values and vectors) or aggregate types,
whose passing can differ largely between architectures.

\uparh{Non-Standard ABIs.}~Customized libraries may employ non-standard or
highly specialized calling conventions 
not covered by our method,
leading to incorrect argument passing or return address handling. In such
cases, the libraries should be translated as well.

While solving these potential issues may require additional expertise and
effort, the root causes can generally be identified by differentially debugging
translated and original binaries. Solutions can then be derived from established
assembly programming practices and ABI documentation.

\section{Implementation}
\label{sec:impl}

The core components of our research prototype, \tool,  consist of roughly 6K
LoC in Python (excluding comments and blank lines), leveraging the \texttt{angr}
symbolic execution engine~\cite{wang2017angr} and GPT-4~\cite{gpt-4} as the
underlying LLM. For disassembling and symbolizing binary code, we chose
\texttt{RetroWrite}, an open-source binary rewriter that has been extensively
evaluated in prior studies ~\cite{dinesh2020retrowrite,kim2023reassembly}.
All binaries were compiled with GCC 11.4.0 on an x86-64 machine
running Ubuntu 22.04, equipped with an Intel Core i7-9700K CPU and 32 GB 
RAM. The translated binaries were executed on a development board
featuring an ARMv8.2-based rk3588
processor~\cite{rk3588}. For comparison with
ExaGear, which is compatible only with
Huawei-manufactured CPUs, we ran translated binaries on a Huawei Cloud virtual
machine powered by a Kunpeng 920 processor (ARMv8.2)~\cite{hisilicon}.

\parh{Target Software.}~Our current implementation targets user-level
single-threaded C programs compiled without anti-analysis techniques like
obfuscation or self-modification. The evaluation mainly covers a subset of
x86-64 ISA, including core general-purpose instructions and frequently used SSE
instructions. Support for floating-point operations is limited, as we currently
exclude x87 FPU instructions. Limitations and future works are further discussed
in \S~\ref{sec:discussion}.

\section{Evaluation}
\label{sec:evaluation}

This section evaluates the performance of \tool\ by answering the
following research questions:

\begin{itemize}[noitemsep,topsep=0pt,leftmargin=3mm]
	\item \textbf{RQ1:} How many translation rules were collected, and
	what were the development costs of \tool?
	\item \textbf{RQ2:} How does the runtime performance of binaries
	translated by \tool compare to existing open-source translators and the
	industrial dynamic binary translator ExaGear?
	\item \textbf{RQ3:} What factors affect the execution efficiency of
	translated binaries?
\end{itemize}

\subsection{Characteristics of Translation Rules}
\label{subsec:evaluation:rules}

To answer \textbf{RQ1}, 
this section details the characteristics of the $3,149$ rules collected by
\tool.
These rules encompass $3,756$
x86-64 instructions and $7,784$ AArch64 instructions in total. On average, each
rule maps $1.2$ x86-64 instructions to $2.5$ AArch64 instructions,
reflecting our observation that converting CISC instructions to RISC
instructions often requires multiple instructions. Meanwhile, as shown in
\T~\ref{tab:rules}, about $95\%$ rules have three or fewer x86-64
instructions, while the number of AArch64 instructions mostly
ranges from one to five. 
Generally, shorter snippets are more likely to match similar
snippets and easier for LLMs to translate. 
While longer snippets likely expose richer optimization opportunities, no
significant correlation was observed between snippet length and LLM failures.
However, it is worth noting that the LLM did not achieve perfect
coverage. Consequently, we manually provided an additional 68 rules for cases
where the LLM failed, the details of which are discussed in \S~\ref{sec:discussion}.

\begin{table}[!htp]
	\centering
	\caption{The number of translation rules of different lengths.} 
	
	\resizebox{0.87\linewidth}{!}{
		\begin{tabular}{l | c | c | c | c | c | c | c | c}
			\hline 
			\textbf{\#Inst} & 1  & 2 & 3 & 4  & 5 & 6 & $\ge7$  & Sum\\ 
			\hline
			\texttt{x86-64} & 2,177 & 524 & 297 & 144 & 4 & 2 & 1 & 3,149 \\	
			
			\texttt{AArch64} & 689 & 677 & 678 & 585 & 279 & 112 & 129 & 3,149 \\	
			\hline
		\end{tabular}
	}
	\label{tab:rules}
	
\end{table}

The rule collection process was highly cost-effective.
The entire process, including iterative corrections, consumed approximately
$2,036K$ tokens, 
totaling less than $\$130$. With an average requirement of under 30
seconds and roughly $\$0.04$ per rule, these results highlight the economic
viability of our approach.
In terms of instruction set coverage, the collected rules cover $148$
unique x86-64 instructions and $102$ AArch64 instructions. While
modern CPUs support a vast number of instructions, many are highly specialized
for specific hardware features (e.g., AVX512 and SGX) or are rarely used.
Translating these hardware-dependent instructions presents 
unique challenges that are outside the scope of this research. 
This work primarily pioneers a new paradigm for binary translation
by focusing on 259 core general-purpose instructions and frequently used SSE
instructions~\cite{dasgupta2019complete,akshintala2019x86,mahoney2021enumerating}.
For
comparison, QEMU contains $283$ code generation functions for
x86-64~\cite{qemu-emit}, each of which translates one x86-64 instruction into
IR. 
The ability of \tool\ to collect a significant set of rules at a very
low cost, covering a substantial portion of the core instruction set, shows high
potential and research value of this approach. 
Interestingly, during our evaluation, we found a program where QEMU failed 
(see \T~\ref{tab:rk3588-1}). QEMU
throws an ``Illegal Instruction'' error when translating the \texttt{shlx \%rsi,
\%rdx, \%rdx} instruction. This suggests that even widely-used translators like
QEMU may still contain flaws, underscoring the motivation
of our work.

\subsection{Performance Overhead}
\label{subsec:evaluation:overhead}

\noindent \textbf{Test Cases.}~We use a comprehensive suite of benchmarks, including
CHStone~\cite{hara2008chstone}, Coremark-Pro~\cite{gal2012exploring,coremark},
and SPEC CPU benchmark~\cite{henning2006spec,bucek2018spec}. We also
include two widely used real-world projects: \texttt{Zydis}~\cite{zydis}, a fast
x86/x64 disassembler library, and \texttt{yyjson}~\cite{yyjson},
a high-performance JSON library. 
Our evaluation encompassed a total of 23 diverse programs, detailed in
\T~\ref{tab:statistic}
and %
\T~\ref{tab:description}. 
This test suite ensures a thorough assessment of \tool's effectiveness across a
range of complex instructions. We empirically consider a translation
correct when the translated binary exhibits consistent input-output behavior on
the test cases.

\begin{table}[h]
	\centering
	\caption{Statistics of evaluated software.
	} 
	
	\resizebox{0.90\linewidth}{!}{

		\begin{tabular}{l | c | c || l | c | c }
			\hline 
			& \textbf{Source}  &  \multirow{2}{*}{\textbf{\#Func}}  &   & \textbf{Source}  & \multirow{2}{*}{\textbf{\#Func}} \\ 
			& \textbf{LoC}  &    &   & \textbf{LoC}  &  \\ 
			\hline
			\texttt{adpcm}& 887 & 15 & \texttt{loops} & 9.5K & 328    \\
			\texttt{gsm}& 634 & 12 & \texttt{parser} & 18K & 310   \\
			\texttt{mips}& 388 & 1 & \texttt{radix2} & 7K & 289   \\
			\texttt{motion}& 1.1K & 14  & \texttt{zlib} & 21.5K & 396   \\
			\texttt{aes}& 1.1K & 11 & \texttt{bzip2} & 8.1K & 70   \\
			\texttt{sha1}& 1.4K & 8 & \texttt{xz} & 35.6K & 390   \\	
			
			\texttt{blowfish}& 1.8K & 6 & \texttt{mcf} & 4K & 43   \\	
   
			\texttt{jpeg}& 3.3K & 29 & \texttt{libquantum} & 4.3K & 98   \\	
   
			\texttt{cjpeg}& 19.5K & 416 & \texttt{hmmer} & 38.3K & 494   \\	
			
			\texttt{sha256}& 7.2K & 283 & \texttt{Zydisn} & 18K & 161   \\	
			
			\texttt{linpack}& 7.5K & 290 & \texttt{yyjson} & 12K & 92   \\	
			\cline{4-6}
			\texttt{nnet}& 8.2K & 288 & Sum & 219.4K & 4,044   \\	
        	\hline
		\end{tabular}
	}
	\label{tab:statistic}
	
\end{table}

\T~\ref{tab:rk3588-1} compares the performance overhead of \tool\ with QEMU, Instrew,
and McSema on the rk3588 platform. Static lifters like Mctoll
and RetDec were excluded because they failed to process any
test case, a limitation consistent with findings in prior
research~\cite{liu2022sok,parzefall2024you,deshpande2024polynima}.
The results reveal that \tool\ consistently and significantly outperforms other
translators.
Notably, McSema failed on 11 of 23 cases, while both Instrew and QEMU
failed on the \texttt{xz} case. These findings underscore substantial
performance and robustness advantages of our approach over
traditional methods.

Moreover, QEMU and Instrew occasionally exhibited substantial overhead (over
20$\times$). Our investigation revealed that these programs' test suites consist
of low-workload cases, causing frequent program restarts. Since translated code
is not retained between runs, each restart forces a repeated translation,
contributing to the high overhead. Moreover, Instrew's reliance on LLVM for
runtime optimization and code generation further adds to this overhead.
Nevertheless, it is important to note that Instrew's novel design is a
significant contribution, offering an efficient and reliable system that
substantially outperforms QEMU in high-workload scenarios. In contrast,
static approaches like McSema and \tool\ turn translation into a one-time
offline effort, producing portable binaries that execute natively. However,
McSema's low success rate highlights the inherent difficulties of static IR
lifting.
We attribute \tool's performance and robustness advantages to its
static assembly-to-assembly translation paradigm, which bypasses the daunting
binary analysis challenges while minimizing runtime overhead.

\begin{table}[!htp]
	\centering
	\caption{
	Performance overhead comparison on
	rk3588 versus native AArch64 execution. Cases with no lifted
	code, compilation failures, or inconsistent results are marked ``failed''.
	} 
	
	\resizebox{0.83\linewidth}{!}{
		\begin{tabular}{l | c | c | c | c }
			\hline 
			\cline{2-5}
			\textbf{Testcase} & \textbf{QEMU}  & \textbf{Instrew}  & \textbf{McSema} & \textbf{\tool}  \\ 
			\hline
			\texttt{adpcm} & 33.57$\times$ & $>$$100\times$ & 1.94$\times$ & \textbf{1.22$\times$}   \\	
			\texttt{gsm} & 33.94$\times$ & $>$$100\times$ & 1.58$\times$ & \textbf{1.23$\times$}   \\	
			\texttt{mips} & 34.08$\times$ & $>$$100\times$ & 1.90$\times$ & \textbf{1.26$\times$}  \\	
			\texttt{motion} & 33.67$\times$ & $>$$100\times$ & 1.87$\times$ & \textbf{1.24$\times$} \\	
			\texttt{aes} & 33.74$\times$ & $>$$100\times$ & 1.84$\times$ & \textbf{1.22$\times$}  \\	
			\texttt{sha1} & 29.84$\times$ & $>$$100\times$ & 2.29$\times$ & \textbf{1.24$\times$}  \\	
			\texttt{blowfish} & 29.16$\times$ & $>$$100\times$ & 2.16$\times$ & \textbf{1.26$\times$}  \\	
			\texttt{jpeg} & 23.03$\times$ & $>$$100\times$ & 3.62$\times$ & \textbf{1.24$\times$}   \\	

			\texttt{cjpeg} & 3.52$\times$ & $>$$100\times$ & failed & \textbf{1.36$\times$}  \\	
			\texttt{sha256} & 32.69$\times$ & $>$$100\times$ & 1.72$\times$ & \textbf{1.19$\times$}  \\	
			\texttt{linpack} & 8.94$\times$ & 90.1$\times$ & failed & \textbf{1.26$\times$}   \\	
			\texttt{nnet} & 15.24$\times$ & 33.3$\times$ & 4.67$\times$ & \textbf{1.86$\times$}   \\	
			\texttt{loops} & 6.51$\times$ & 76.4$\times$ & failed & \textbf{1.27$\times$}  \\	
			\texttt{parser} & 2.97$\times$ & 83.7$\times$ & 1.88$\times$ & \textbf{1.03$\times$}   \\	
			\texttt{radix2} & 25.58$\times$ & $>$$100\times$ & 46.04$\times$ & \textbf{1.27$\times$}  \\	
			\texttt{zlib} & 17.61$\times$ & $>$$100\times$ & failed & \textbf{1.15$\times$}   \\	

			\texttt{bzip2} & 12.82$\times$ & 1.32$\times$ & failed & \textbf{1.14$\times$}    \\	
			\texttt{xz} & failed & failed & failed & \textbf{1.25$\times$}   \\				
			\texttt{mcf} & 3.89$\times$ & 1.49$\times$ & failed & \textbf{1.13$\times$}   \\	
			\texttt{libquantum} & 5.55$\times$ & 4.62$\times$ & failed & \textbf{1.12$\times$}   \\	
			\texttt{hmmer} & 5.69$\times$ & 3.03$\times$ & failed & \textbf{1.39$\times$}  \\	
   
   			\texttt{Zydis} & 21.34$\times$ & $>$$100\times$ & failed & \textbf{1.16$\times$}   \\	
			
   			\texttt{yyjson} & 25.36$\times$ & $>$$100\times$ & failed & \textbf{1.44$\times$}   \\	
            \hline 
		\end{tabular}
    }
	\label{tab:rk3588-1}
	
\end{table}

\T~\ref{tab:kunpeng} further compares \tool's performance against ExaGear. 
While both tools exhibit strengths,
\tool\ demonstrates consistently stable and superior performance overall.
Similarly, 
ExaGear's performance suffers from significant overhead in low-workload scenarios.
Given that ExaGear is a mature, heavily optimized
commercial product developed by a leading chip
manufacturer, these results highlight the remarkable potential and research
value of our approach.

\begin{table}[h]
	\centering
	\caption{Performance overhead of ExaGear and \tool\ on 
	Kunpeng 920 versus native AArch64 execution.
	} 
	
	\resizebox{0.99\linewidth}{!}{

		\begin{tabular}{l | c | c || l | c | c }
			\hline 
			\textbf{Testcases} & \textbf{ExaGear}  &   \textbf{\tool} &  \textbf{Testcases} & \textbf{ExaGear}  & \textbf{\tool} \\ 
			\hline
			\texttt{adpcm}& 45.36$\times$ & \textbf{1.12$\times$} & \texttt{loops} & 4.33$\times$ & \textbf{1.94$\times$}    \\
			\texttt{gsm}& 47.90$\times$ & \textbf{1.16$\times$} & \texttt{parser} & 2.50$\times$ & \textbf{1.09$\times$}   \\
			\texttt{mips}& 47.03$\times$ & \textbf{1.17$\times$} & \texttt{radix2} & 36.67$\times$ & \textbf{1.19$\times$}   \\
			\texttt{motion}& 47.98$\times$ & \textbf{1.22$\times$}  & \texttt{zlib} & 24.54$\times$ & \textbf{1.10$\times$}   \\
			\texttt{aes}& 48.45$\times$ & \textbf{1.21$\times$} & \texttt{bzip2} & \textbf{1.09$\times$} & 1.25$\times$   \\
			\texttt{sha1}& 44.11$\times$ & \textbf{1.24$\times$} & \texttt{xz} & \textbf{1.10$\times$} & 1.23$\times$   \\	
			
			\texttt{blowfish}& 41.11$\times$ & \textbf{1.27$\times$} & \texttt{mcf} & 1.21$\times$ & \textbf{1.13$\times$}   \\	
   
			\texttt{jpeg}& 30.63$\times$ & \textbf{1.18$\times$} & \texttt{libquantum} & \textbf{1.53$\times$} & 1.69$\times$   \\	
   
			\texttt{cjpeg}& 4.80$\times$ & \textbf{1.37$\times$} & \texttt{hmmer} & \textbf{1.20$\times$} & 1.24$\times$   \\	
			
			\texttt{sha256}& 46.88$\times$ & \textbf{1.20$\times$} & \texttt{Zydisn} & 37.90$\times$ & \textbf{1.08$\times$}   \\	
			
			\texttt{linpack}& 2.30$\times$ & \textbf{1.19$\times$} & \texttt{yyjson} & 28.43$\times$ & \textbf{1.59$\times$}   \\	
			\texttt{nnet}& \textbf{2.13$\times$} & 2.44$\times$ &  &  &    \\	
            \hline
		\end{tabular}
	}
	\label{tab:kunpeng}
	
\end{table}

\subsection{Performance Analysis}
\label{subsec:evaluation:algorithms}

To better understand why \tool's performance varies across different cases, we
conducted detailed profiling on three typical algorithms from a popular C
library~\cite{c-algo}: \texttt{avl-tree}, \texttt{binary-heap}, and
\texttt{hash-table}. 
Using the \texttt{perf} profiling tool on the rk3588 platform, \T~\ref{tab:algo}
presents the number of executed instructions and CPU
cycles. 

\begin{table}[!htp]
	\centering
	\caption{Evaluation results of three algorithms. \texttt{N} and \texttt{T} represent the native binary and translated binary, respectively.}

	\resizebox{0.98\linewidth}{!}{
		\begin{tabular}{l | c | c | c | c}
			\hline 
			\multirow{2}{*}{\textbf{Algorithm}} & \textbf{Source}  &   \textbf{\#Executed} &   \textbf{\#CPU Cycles} &   \multirow{2}{*}{\textbf{Performance}}  \\ 
			& \textbf{LoC}  &  \textbf{Insts} (\texttt{N} / \texttt{T}) &  (\texttt{N} / \texttt{T})  &   \textbf{}  \\ 
			\hline
			\texttt{avl-tree} & 1.2K & 2,327M / 3,526M & 1,016M / 1,292M  & 1.20$\times$ \\	
			
			\texttt{binary-heap} & 0.5K & 143.7M / 208M & 120.4M / 120.8M  & 1.02$\times$ \\	
   
			\texttt{hash-table} & 1.1K & 353.3M / 477.4M & 210M / 225M  & 1.12$\times$ \\	
   
   			\hline
		\end{tabular}
	}
	\label{tab:algo}

\end{table}

As shown in \T~\ref{tab:algo}, translated binaries (marked \texttt{T})
consistently execute more instructions and consume more CPU cycles than native
binaries (marked \texttt{N}). However, the increase in cycles is not always
proportional to the increase in instructions. For instance, while the translated
\texttt{avl-tree} executes 50\% more instructions, it consumes only 27\% more
CPU cycles. This discrepancy suggests that performance is heavily influenced by
microarchitectural features, such as pipelining, branch prediction, and
caching. 
The instruction sequence and memory access patterns in
translated binaries differ considerably from those in compiler-generated native
code, which can substantially affect execution speed.
Our evaluation indicates that the CPU pipeline effectively mitigates
much of the overhead from increased instruction count.

\begin{table}[!htp]
	\centering
	\caption{Detailed performance profiling with \texttt{perf}. The data is collected on the rk3588 device.} 

	\resizebox{0.99\linewidth}{!}{
		\begin{tabular}{l | c | c | c | c}
			\hline 
			\multirow{3}{*}{\textbf{Algorithm}} & \textbf{Branch}  &   \textbf{Cache} &   \textbf{Stalled Cycles} &   \textbf{Stalled Cycles}  \\ 
			& \textbf{Misses}  &  \textbf{Misses}  &  \textbf{Frontend}  &   \textbf{Backend} 
			 \\ 
			 & (\texttt{N} / \texttt{T})  &  (\texttt{N} / \texttt{T})  &  (\texttt{N} / \texttt{T})  &   (\texttt{N} / \texttt{T}) \\  
			 \hline
			\texttt{avl-tree} & 2.3M / 2.1M & 12M / 12.2M & 123.4M / 10.6M  & 31.8M / 131.6M \\	
			
			\texttt{binary-heap} & 135K / 126.6K & 32K / 34K & 332K / 334K  & 28.5M / 25M \\	
   
			\texttt{hash-table} & 61.2K / 56K & 299K / 316K & 1.8M / 1.9M  & 42M / 41.7M \\	
   
   			\hline
		\end{tabular}
	}
	\label{tab:perf}

\end{table}

For a deeper insight, we used \texttt{perf} to record
microarchitecture level metrics, with the results shown in Table \ref{tab:perf}.
The data reveals a key finding: 
native binaries generally exhibit more branch misses, whereas translated
binaries suffer from more cache misses.  
This is likely because the repetitive, patterned code resulting from translation
rules is more friendly to branch predictors. 
However, the translation process does not account
for memory layout, leading to lower cache efficiency. 
The \texttt{avl-tree} is a clear example, with a significant increase in cache
misses (over 200K more), which results in a high number of backend
stalled cycles (notably, one cache miss can cause a long stall). 
This indicates that \texttt{avl-tree}'s overhead
stems primarily from poor cache efficiency. 

Overall, executing more instructions does not necessarily lead to
high overhead. 
Performance depends heavily on how well the translated code utilizes CPU
microarchitecture.
Designing a binary translator that effectively accounts for
the microarchitecture remains an open challenge,
and investigating the
intricate relationship between binary code and CPU pipeline behavior is an
intriguing topic for future research.

\section{Discussion}
\label{sec:discussion}

\parh{Remaining Challenges.}~While this work establishes the viability
of a novel, LLM-based static translation pipeline, several orthogonal challenges remain.
These long-standing issues represent important directions
for future work, each requiring
substantial, dedicated research and engineering efforts beyond the scope of a
single paper.

\sparh{1) Concurrent Programs.}~Translating concurrent programs is
challenging due to the different memory models
between architectures.
Nevertheless, prior research offers a solid foundation by inserting and
optimizing fences in lifted IR
code~\cite{rocha2022lasagne,gouicem2022risotto,deshpande2024polynima,beck2023atomig}.
Although these methods do not focus on assembly-level translation, future
research could adapt them to our static assembly-to-assembly translation
paradigm.

\sparh{2) C++ Programs.}~C++ features like exceptions and virtual functions,
which rely on compiler-specific implementations, present unique hurdles for
binary rewriting. These features complicate CFGs and data flow, making
disassembly more difficult. Since our method relies on correct disassembly
code, future research could focus on improving existing binary rewriting and
disassembly techniques to cover these complex language features.

\sparh{3) Floating-Point Operations.}~Architectural differences complicate the
translation of floating-point operations.
The x86-64 uses an eight-register stack for its x87 Floating Point Unit
(FPU)~\cite{kusswurm2014x87}, 
where AArch64's Advanced SIMD (NEON)~\cite{neon,lee2018efficient} lacks a
floating-point stack.
Translating these hardware-dependent instructions demands significant manual
effort and expertise that is beyond this paper's scope. 
Current solutions often rely on carefully crafted 
helper functions. Due to this complexity,
we exclude these instructions from our implementation,
leaving a more efficient automated method as future work.

\sparh{4) System Calls.}~
Our evaluation targets user-level applications where dynamically linked native
libraries handle system calls, simplifying translation since library APIs
are consistent across architectures.
However, translating system calls is inherently complex due to differences in
conventions and data structure layouts. Our preliminary findings reveal
inconsistencies in binary-level data structures between Linux kernels of
different architectures (more details in the
Appendix~\ref{subsec:appendix:stat}). 
Initial experiments suggest these issues can be resolved
by patching native libraries to enforce data structure consistency, with minimal
performance impact. 
Future research could explore automated methods to
identify and resolve such issues. 

\noindent \textbf{Supplementing with Manually-Crafted Rules.}~Although LLMs show
significant potential in binary translation,
they can struggle with complex instructions due to reasoning limitations.
While our symbolic verification method automatically detects such failures, 
fixing them may require human intervention. 
In our experiments, we manually created 68 translation rules for cases in which LLM
failed.

The practicality of this manual approach is supported by our methodology, which
intentionally keeps translation code snippets short and their data flow simple.
This design allows an experienced developer, referencing the relevant ARM
documentation, to author an equivalent AArch64 code for a given snippet in
minutes without extensive coding. While this requires expertise, our automatic
semantic verification drastically reduces the manual validation effort,
accelerating the entire workflow. Once verified, the translated code is
parameterized into a reusable rule.

\section{Related Work}
\label{sec:related}

\parh{Translation Validation.}~Ensuring the correctness of code transformations
is a critical challenge. Recent efforts~\cite{dasgupta2020scalable} have focused
on validating the process of lifting binaries to LLVM IR by employing scalable
instruction-level checking and formal verification to guarantee semantic
equivalence. Within the compiler toolchain, tools like
Alive2~\cite{lopes2021alive2} provide bounded translation validation to find
bugs in LLVM's optimization passes. Other techniques, such as semantic program
alignment~\cite{churchill2019semantic}, verify equivalence by matching program
states along their execution traces based on semantic information, a method
particularly effective for complex transformations that significantly
restructure code. Complementing these, language-parametric
approaches~\cite{kasampalis2021language} offer more generalized validation
frameworks that are not tied to a specific transformation, further strengthening
the reliability of modern compilers.
Nevertheless, these works primarily target high-level IR transformations,
whereas our work is centered on assembly-level cross-architecture translation. 

\parh{Extensions in Binary Translation.}~Recent advancements in binary
translation have focused on addressing the complexities of modern software,
particularly in areas like concurrency and post-compilation optimization. The
challenge of translating for weak memory model architectures has been tackled by
both static approaches, like Lasagne~\cite{rocha2022lasagne}, and dynamic systems
such as Risotto~\cite{gouicem2022risotto}. For multithreaded applications,
hybrid recompilation techniques like Polynima~\cite{deshpande2024polynima} have
emerged to balance performance and translation overhead. 
Researchers have also developed solutions for specific technical hurdles that
impact correctness and efficiency. For instance, WYTIWYG~\cite{parzefall2024you}
presents a method for dynamic stack-layout recovery, which enables
post-compilation optimization for binaries, while
Tiaozhuan~\cite{li2025tiaozhuan} introduces a general optimization for indirect
branches, a common performance bottleneck. Another promising direction involves
leveraging information beyond the raw binary. Plankton~\cite{zhou2024plankton},
for example, supports full-fledged static analysis on lifted IR by utilizing
debug information, while Biotite~\cite{chen2025biotite} uses source-level
knowledge to generate higher-performance code.
In the broader context of binary lifting, adjacent research explores
decompilation to recover high-level source code (e.g., C) from binaries.
Although decompilation is inherently challenging and often
error-prone~\cite{liu2020far}, recent advancements utilizing LLMs have
substantially improved decompilation fidelity for small-scale
programs~\cite{tan2024llm4decompile,wong2025decllm,zhou2026fidelitygpt}.
They are, however, orthogonal to our work, which focuses on binary-level
translation that produces efficient code and supports scalable verification.

\parh{Automated Generation of Translation Logic.}~A significant line of research
has focused on automating the creation of translation logic to reduce manual
effort and improve the quality of generated code. Early work in this area
utilized peephole superoptimizers~\cite{bansal2008binary} to automatically
discover optimal, short instruction sequences for translation. 
Other research has improved the front-end of the translation process, presenting
novel compiler-leveraged techniques for lifting assembly to a high-level
intermediate representation~\cite{hasabnis2016lifting}.
More recently, this has evolved into a broader learning-based paradigm. For
instance, several studies have shown how translation rules for dynamic binary
translators can be automatically learned from compiler-generated binary code, an
approach demonstrated by Wang et al.~\cite{wang2018enhancing} and Song et
al.~\cite{song2019unleashing}. This learning-based methodology has been
successfully extended to the more demanding context of system-level
translation~\cite{jiang2024system} and has been made more data-efficient through
techniques like parameterization~\cite{jiang2020more}, which allows 
deriving more general rules from less training data. 
In contrast, our method leverages LLMs to directly generate high-quality
assembly code, eliminating the reliance on extensive training data or a separate
compilation process.

\section{Conclusion}
\label{sec:conclusion}

This paper presents a novel static binary translation paradigm that converts
x86-64 binaries to AArch64,
producing
portable binaries with near native performance. By utilizing LLMs, our approach
substantially reduces the manual effort and
expertise required to build translation engines.
We ensure translation correctness through efficient semantic verification of
small, simplified code snippets.
Our research prototype, \tool, has been shown to outperform both traditional
open-source tools and a leading industrial translator.
This work signals a promising new avenue for research in binary translation
techniques.

\section*{Acknowledgments}
This work was supported in part by the Hong Kong RGC Postdoctoral Fellowship
Scheme (PDFS2324-6S08), the HKUST Bridge-the-Gap Fund (BGF.001.2025), and the
Fundamental and Interdisciplinary Disciplines Breakthrough Plan of the Ministry
of Education of China (No. JYB2025XDXM118). We also thank the HKUST Fok Ying
Tung Research Institute and the National Supercomputing Center in Guangzhou
(Nansha Sub-center) for providing computational resources, and the Collaborative
Innovation Center of Novel Software Technology and Industrialization (Jiangsu,
China) for their support.

\bibliographystyle{plain}
\bibliography{bib/references,bib/llm,bib/binary}

@inproceedings{de2000static,
  title={On the static analysis of indirect control transfers in binaries},
  author={De Sutter, B and De Bosschere, K and Keyngnaert, Peter and Demoen, Bart},
  booktitle={Proceedings of the International Conference on Parallel and Distributed Processing Techniques and Applications},
  volume={2},
  pages={1013--1019},
  year={2000},
  organization={CSREA PRESS}
}

@article{kinder2010static,
  title={Static analysis of x86 executables},
  author={Kinder, Johannes},
  year={2010},
  publisher={Technische Universit{\"a}t Darmstadt}
}

@inproceedings{yu2022deepdi,
  title={$\{$DeepDi$\}$: Learning a relational graph convolutional network model on instructions for fast and accurate disassembly},
  author={Yu, Sheng and Qu, Yu and Hu, Xunchao and Yin, Heng},
  booktitle={31st USENIX Security Symposium (USENIX Security 22)},
  pages={2709--2725},
  year={2022}
}

@article{pei2020xda,
  title={Xda: Accurate, robust disassembly with transfer learning},
  author={Pei, Kexin and Guan, Jonas and Williams-King, David and Yang, Junfeng and Jana, Suman},
  journal={arXiv preprint arXiv:2010.00770},
  year={2020}
}

@inproceedings{ye2023d,
  title={D-arm: Disassembling arm binaries by lightweight superset instruction interpretation and graph modeling},
  author={Ye, Yapeng and Zhang, Zhuo and Shi, Qingkai and Aafer, Yousra and Zhang, Xiangyu},
  booktitle={2023 IEEE Symposium on Security and Privacy (SP)},
  pages={2391--2408},
  year={2023},
  organization={IEEE}
}

@inproceedings{andriesse2016depth,
  title={An $\{$In-Depth$\}$ Analysis of Disassembly on $\{$Full-Scale$\}$ x86/x64 Binaries},
  author={Andriesse, Dennis and Chen, Xi and Van Der Veen, Victor and Slowinska, Asia and Bos, Herbert},
  booktitle={25th USENIX security symposium (USENIX security 16)},
  pages={583--600},
  year={2016}
}

@inproceedings{kim2023reassembly,
  title={Reassembly is hard: a reflection on challenges and strategies},
  author={Kim, Hyungseok and Kim, Soomin and Lee, Junoh and Jee, Kangkook and Cha, Sang Kil},
  booktitle={32nd USENIX Security Symposium (USENIX Security 23)},
  pages={1469--1486},
  year={2023}
}

@inproceedings{wang2015reassembleable,
  title={Reassembleable disassembling},
  author={Wang, Shuai and Wang, Pei and Wu, Dinghao},
  booktitle={24th USENIX Security Symposium (USENIX Security 15)},
  pages={627--642},
  year={2015}
}

@inproceedings{wang2017ramblr,
  title={Ramblr: Making Reassembly Great Again.},
  author={Wang, Ruoyu and Shoshitaishvili, Yan and Bianchi, Antonio and Machiry, Aravind and Grosen, John and Grosen, Paul and Kruegel, Christopher and Vigna, Giovanni},
  booktitle={NDSS},
  year={2017}
}

@inproceedings{flores2020datalog,
  title={Datalog disassembly},
  author={Flores-Montoya, Antonio and Schulte, Eric},
  booktitle={29th USENIX Security Symposium (USENIX Security 20)},
  pages={1075--1092},
  year={2020}
}

@inproceedings{dinesh2020retrowrite,
  title={Retrowrite: Statically instrumenting cots binaries for fuzzing and sanitization},
  author={Dinesh, Sushant and Burow, Nathan and Xu, Dongyan and Payer, Mathias},
  booktitle={2020 IEEE Symposium on Security and Privacy (SP)},
  pages={1497--1511},
  year={2020},
  organization={IEEE}
}

@inproceedings{williams2020egalito,
  title={Egalito: Layout-agnostic binary recompilation},
  author={Williams-King, David and Kobayashi, Hidenori and Williams-King, Kent and Patterson, Graham and Spano, Frank and Wu, Yu Jian and Yang, Junfeng and Kemerlis, Vasileios P},
  booktitle={Proceedings of the Twenty-Fifth International Conference on Architectural Support for Programming Languages and Operating Systems},
  pages={133--147},
  year={2020}
}

@inproceedings{cadar2008klee,
  title={Klee: unassisted and automatic generation of high-coverage tests for complex systems programs.},
  author={Cadar, Cristian and Dunbar, Daniel and Engler, Dawson R and others},
  booktitle={OSDI},
  volume={8},
  pages={209--224},
  year={2008}
}

@article{baldoni2018survey,
  title={A survey of symbolic execution techniques},
  author={Baldoni, Roberto and Coppa, Emilio and D’elia, Daniele Cono and Demetrescu, Camil and Finocchi, Irene},
  journal={ACM Computing Surveys (CSUR)},
  volume={51},
  number={3},
  pages={1--39},
  year={2018},
  publisher={ACM New York, NY, USA}
}

@inproceedings{de2008z3,
  title={Z3: An efficient SMT solver},
  author={De Moura, Leonardo and Bj{\o}rner, Nikolaj},
  booktitle={International conference on Tools and Algorithms for the Construction and Analysis of Systems},
  pages={337--340},
  year={2008},
  organization={Springer}
}

@article{li2014symbolic,
  title={Symbolic optimization with SMT solvers},
  author={Li, Yi and Albarghouthi, Aws and Kincaid, Zachary and Gurfinkel, Arie and Chechik, Marsha},
  journal={ACM SIGPLAN Notices},
  volume={49},
  number={1},
  pages={607--618},
  year={2014},
  publisher={ACM New York, NY, USA}
}

@inproceedings{poeplau2020symbolic,
  title={Symbolic execution with $\{$SymCC$\}$: Don't interpret, compile!},
  author={Poeplau, Sebastian and Francillon, Aur{\'e}lien},
  booktitle={29th USENIX Security Symposium (USENIX Security 20)},
  pages={181--198},
  year={2020}
}

@inproceedings{qi2024symfit,
  title={$\{$SymFit$\}$: Making the Common (Concrete) Case Fast for $\{$Binary-Code$\}$ Concolic Execution},
  author={Qi, Zhenxiao and Hu, Jie and Xiao, Zhaoqi and Yin, Heng},
  booktitle={33rd USENIX Security Symposium (USENIX Security 24)},
  pages={415--432},
  year={2024}
}

@inproceedings{poeplau2021symqemu,
  title={SymQEMU: Compilation-based symbolic execution for binaries},
  author={Poeplau, Sebastian and Francillon, Aur{\'e}lien},
  booktitle={NDSS 2021, Network and Distributed System Security Symposium},
  year={2021},
  organization={Internet Society}
}

@inproceedings{wang2017angr,
  title={Angr-the next generation of binary analysis},
  author={Wang, Fish and Shoshitaishvili, Yan},
  booktitle={2017 IEEE Cybersecurity Development (SecDev)},
  pages={8--9},
  year={2017},
  organization={IEEE}
}

@inproceedings{mahoney2021enumerating,
  title={Enumerating x86-64 - it's not as easy as counting},
  author={Mahoney, William and McDonald, J Todd},
  booktitle={Technical report, University of Nebraska Omaha},
  year={2021}
}

@inproceedings{dasgupta2019complete,
  title={A complete formal semantics of x86-64 user-level instruction set architecture},
  author={Dasgupta, Sandeep and Park, Daejun and Kasampalis, Theodoros and Adve, Vikram S and Ro{\c{s}}u, Grigore},
  booktitle={Proceedings of the 40th ACM SIGPLAN Conference on Programming Language Design and Implementation},
  pages={1133--1148},
  year={2019}
}

@inproceedings{akshintala2019x86,
  title={X86-64 instruction usage among c/c++ applications},
  author={Akshintala, Amogh and Jain, Bhushan and Tsai, Chia-Che and Ferdman, Michael and Porter, Donald E},
  booktitle={Proceedings of the 12th ACM International Conference on Systems and Storage},
  pages={68--79},
  year={2019}
}

@inproceedings{kim2025towards,
  title={Towards Sound Reassembly of Modern x86-64 Binaries},
  author={Kim, Hyungseok and Kim, Soomin and Cha, Sang Kil},
  booktitle={Proceedings of the 30th ACM International Conference on Architectural Support for Programming Languages and Operating Systems, Volume 2},
  pages={1317--1333},
  year={2025}
}

@inproceedings{liu2022sok,
  title={Sok: Demystifying binary lifters through the lens of downstream applications},
  author={Liu, Zhibo and Yuan, Yuanyuan and Wang, Shuai and Bao, Yuyan},
  booktitle={2022 IEEE Symposium on Security and Privacy (SP)},
  pages={1100--1119},
  year={2022},
  organization={IEEE}
}

@inproceedings{parzefall2024you,
  title={What you trace is what you get: dynamic stack-layout recovery for binary recompilation},
  author={Parzefall, Fabian and Deshpande, Chinmay and Hetzelt, Felicitas and Franz, Michael},
  booktitle={Proceedings of the 29th ACM International Conference on Architectural Support for Programming Languages and Operating Systems, Volume 2},
  pages={1250--1263},
  year={2024}
}

@inproceedings{yadavalli2019raising,
  title={Raising binaries to llvm ir with mctoll (wip paper)},
  author={Yadavalli, S Bharadwaj and Smith, Aaron},
  booktitle={Proceedings of the 20th ACM SIGPLAN/SIGBED International Conference on Languages, Compilers, and Tools for Embedded Systems},
  pages={213--218},
  year={2019}
}

@misc{retdec,
  title={Retargetable machine-code decompiler based on LLVM.},
  author={Avast},
  howpublished = "\url{https://github.com/avast/retdec}",
  year = {2025},
}

@inproceedings{di2017rev,
  title={rev. ng: a unified binary analysis framework to recover CFGs and function boundaries},
  author={Di Federico, Alessandro and Payer, Mathias and Agosta, Giovanni},
  booktitle={Proceedings of the 26th International Conference on Compiler Construction},
  pages={131--141},
  year={2017}
}

@inproceedings{di2018rev,
  title={rev. ng: A multi-architecture framework for reverse engineering and vulnerability discovery},
  author={Di Federico, Alessandro and Fezzardi, Pietro and Agosta, Giovanni},
  booktitle={2018 International Carnahan Conference on Security Technology (ICCST)},
  pages={1--5},
  year={2018},
  organization={IEEE}
}

@inproceedings{rocha2022lasagne,
  title={Lasagne: A static binary translator for weak memory model architectures},
  author={Rocha, Rodrigo CO and Sprokholt, Dennis and Fink, Martin and Gouicem, Redha and Spink, Tom and Chakraborty, Soham and Bhatotia, Pramod},
  booktitle={Proceedings of the 43rd ACM SIGPLAN International Conference on Programming Language Design and Implementation},
  pages={888--902},
  year={2022}
}

@inproceedings{meng2016binary,
  title={Binary code is not easy},
  author={Meng, Xiaozhu and Miller, Barton P},
  booktitle={Proceedings of the 25th International Symposium on Software Testing and Analysis},
  pages={24--35},
  year={2016}
}

@inproceedings{shoshitaishvili2016sok,
  title={Sok:(state of) the art of war: Offensive techniques in binary analysis},
  author={Shoshitaishvili, Yan and Wang, Ruoyu and Salls, Christopher and Stephens, Nick and Polino, Mario and Dutcher, Andrew and Grosen, John and Feng, Siji and Hauser, Christophe and Kruegel, Christopher and others},
  booktitle={2016 IEEE symposium on security and privacy (SP)},
  pages={138--157},
  year={2016},
  organization={IEEE}
}

@inproceedings{he2018debin,
  title={Debin: Predicting debug information in stripped binaries},
  author={He, Jingxuan and Ivanov, Pesho and Tsankov, Petar and Raychev, Veselin and Vechev, Martin},
  booktitle={Proceedings of the 2018 ACM SIGSAC Conference on Computer and Communications Security},
  pages={1667--1680},
  year={2018}
}

@inproceedings{xie2024resym,
  title={Resym: Harnessing llms to recover variable and data structure symbols from stripped binaries},
  author={Xie, Danning and Zhang, Zhuo and Jiang, Nan and Xu, Xiangzhe and Tan, Lin and Zhang, Xiangyu},
  booktitle={Proceedings of the 2024 on ACM SIGSAC Conference on Computer and Communications Security},
  pages={4554--4568},
  year={2024}
}

@inproceedings{marcelli2022machine,
  title={How machine learning is solving the binary function similarity problem},
  author={Marcelli, Andrea and Graziano, Mariano and Ugarte-Pedrero, Xabier and Fratantonio, Yanick and Mansouri, Mohamad and Balzarotti, Davide},
  booktitle={31st USENIX Security Symposium (USENIX Security 22)},
  pages={2099--2116},
  year={2022}
}

@inproceedings{mantovani2022convergence,
  title={The Convergence of Source Code and Binary Vulnerability Discovery--A Case Study},
  author={Mantovani, Alessandro and Compagna, Luca and Shoshitaishvili, Yan and Balzarotti, Davide},
  booktitle={Proceedings of the 2022 ACM on Asia Conference on Computer and Communications Security},
  pages={602--615},
  year={2022}
}

@inproceedings{dinaburg2014mcsema,
  title={Mcsema: Static translation of x86 instructions to llvm},
  author={Dinaburg, Artem and Ruef, Andrew},
  booktitle={ReCon 2014 Conference, Montreal, Canada},
  year={2014}
}

@misc{mcsema,
  title={Framework for lifting x86, amd64, aarch64, sparc32, and sparc64 program binaries to LLVM bitcode.},
  author={Trail of Bits},
  howpublished = "\url{https://github.com/lifting-bits/mcsema}",
  year = {2025},
}

@misc{ida,
  title={IDA Pro  powerful disassembler, decompiler and a versatile debugger.},
  author={Hex-Rays},
  howpublished = "\url{https://hex-rays.com/ida-pro}",
  year = {2025},
}

@inproceedings{engelke2020instrew,
  title={Instrew: Leveraging LLVM for high performance dynamic binary instrumentation},
  author={Engelke, Alexis and Schulz, Martin},
  booktitle={Proceedings of the 16th ACM SIGPLAN/SIGOPS International Conference on Virtual Execution Environments},
  pages={172--184},
  year={2020}
}

@inproceedings{altinay2020binrec,
  title={BinRec: dynamic binary lifting and recompilation},
  author={Altinay, Anil and Nash, Joseph and Kroes, Taddeus and Rajasekaran, Prabhu and Zhou, Dixin and Dabrowski, Adrian and Gens, David and Na, Yeoul and Volckaert, Stijn and Giuffrida, Cristiano and others},
  booktitle={Proceedings of the Fifteenth European Conference on Computer Systems},
  pages={1--16},
  year={2020}
}

@inproceedings{deshpande2024polynima,
  title={Polynima: Practical hybrid recompilation for multithreaded binaries},
  author={Deshpande, Chinmay and Parzefall, Fabian and Hetzelt, Felicitas and Franz, Michael},
  booktitle={Proceedings of the Nineteenth European Conference on Computer Systems},
  pages={1126--1141},
  year={2024}
}

@inproceedings{gouicem2022risotto,
  title={Risotto: A dynamic binary translator for weak memory model architectures},
  author={Gouicem, Redha and Sprokholt, Dennis and Ruehl, Jasper and Rocha, Rodrigo CO and Spink, Tom and Chakraborty, Soham and Bhatotia, Pramod},
  booktitle={Proceedings of the 28th ACM International Conference on Architectural Support for Programming Languages and Operating Systems, Volume 1},
  pages={107--122},
  year={2022}
}

@misc{revng-issue,
  title={Failure when trying to translate a simple dynamic x86-64 binary compiled with clang.},
  author={Rev.ng team},
  howpublished = "\url{https://github.com/revng/revng/issues/317#issuecomment-1721136071}",
  year = {2025},
}

@inproceedings{schwartz2010all,
  title={All you ever wanted to know about dynamic taint analysis and forward symbolic execution (but might have been afraid to ask)},
  author={Schwartz, Edward J and Avgerinos, Thanassis and Brumley, David},
  booktitle={2010 IEEE symposium on Security and privacy},
  pages={317--331},
  year={2010},
  organization={IEEE}
}

@inproceedings{dasgupta2020scalable,
  title={Scalable validation of binary lifters},
  author={Dasgupta, Sandeep and Dinesh, Sushant and Venkatesh, Deepan and Adve, Vikram S and Fletcher, Christopher W},
  booktitle={Proceedings of the 41st ACM SIGPLAN Conference on Programming Language Design and Implementation},
  pages={655--671},
  year={2020}
}

@inproceedings{lopes2021alive2,
  title={Alive2: bounded translation validation for LLVM},
  author={Lopes, Nuno P and Lee, Juneyoung and Hur, Chung-Kil and Liu, Zhengyang and Regehr, John},
  booktitle={Proceedings of the 42nd ACM SIGPLAN International Conference on Programming Language Design and Implementation},
  pages={65--79},
  year={2021}
}

@inproceedings{churchill2019semantic,
  title={Semantic program alignment for equivalence checking},
  author={Churchill, Berkeley and Padon, Oded and Sharma, Rahul and Aiken, Alex},
  booktitle={Proceedings of the 40th ACM SIGPLAN Conference on Programming Language Design and Implementation},
  pages={1027--1040},
  year={2019}
}

@inproceedings{kasampalis2021language,
  title={Language-parametric compiler validation with application to LLVM},
  author={Kasampalis, Theodoros and Park, Daejun and Lin, Zhengyao and Adve, Vikram S and Ro{\c{s}}u, Grigore},
  booktitle={Proceedings of the 26th ACM International Conference on Architectural Support for Programming Languages and Operating Systems},
  pages={1004--1019},
  year={2021}
}

@article{li2025tiaozhuan,
  title={Tiaozhuan: A General and Efficient Indirect Branch Optimization for Binary Translation},
  author={Li, Xinyu and Guo, Guangyao and Lan, Yanzhi and Xue, Feng and Han, Chenji and Niu, Gen and Zhang, Fuxin},
  journal={ACM Transactions on Architecture and Code Optimization},
  volume={22},
  number={1},
  pages={1--26},
  year={2025},
  publisher={ACM New York, NY}
}

@inproceedings{zhou2024plankton,
  title={Plankton: Reconciling binary code and debug information},
  author={Zhou, Anshunkang and Ye, Chengfeng and Huang, Heqing and Cai, Yuandao and Zhang, Charles},
  booktitle={Proceedings of the 29th ACM International Conference on Architectural Support for Programming Languages and Operating Systems, Volume 2},
  pages={912--928},
  year={2024}
}

@inproceedings{chen2025biotite,
  title={Biotite: A High-Performance Static Binary Translator using Source-Level Information},
  author={Chen, Changbin and Sugita, Shu and Nada, Yotaro and Irie, Hidetsugu and Sakai, Shuichi and Shioya, Ryota},
  booktitle={Proceedings of the 34th ACM SIGPLAN International Conference on Compiler Construction},
  pages={167--179},
  year={2025}
}

@inproceedings{bansal2008binary,
  title={Binary Translation Using Peephole Superoptimizers.},
  author={Bansal, Sorav and Aiken, Alex},
  booktitle={OSDI},
  volume={8},
  pages={177--192},
  year={2008}
}

@inproceedings{hasabnis2016lifting,
  title={Lifting assembly to intermediate representation: A novel approach leveraging compilers},
  author={Hasabnis, Niranjan and Sekar, R},
  booktitle={Proceedings of the Twenty-First International Conference on Architectural Support for Programming Languages and Operating Systems},
  pages={311--324},
  year={2016}
}

@inproceedings{liu2020far,
  title={How far we have come: Testing decompilation correctness of C decompilers},
  author={Liu, Zhibo and Wang, Shuai},
  booktitle={Proceedings of the 29th ACM SIGSOFT International Symposium on Software Testing and Analysis},
  pages={475--487},
  year={2020}
}

@article{wong2025decllm,
  title={Decllm: Llm-augmented recompilable decompilation for enabling programmatic use of decompiled code},
  author={Wong, Wai Kin and Wu, Daoyuan and Wang, Huaijin and Li, Zongjie and Liu, Zhibo and Wang, Shuai and Tang, Qiyi and Nie, Sen and Wu, Shi},
  journal={Proceedings of the ACM on Software Engineering},
  volume={2},
  number={ISSTA},
  pages={1841--1864},
  year={2025},
  publisher={ACM New York, NY, USA}
}

@inproceedings{tan2024llm4decompile,
  title={Llm4decompile: Decompiling binary code with large language models},
  author={Tan, Hanzhuo and Luo, Qi and Li, Jing and Zhang, Yuqun},
  booktitle={Proceedings of the 2024 Conference on Empirical Methods in Natural Language Processing},
  pages={3473--3487},
  year={2024}
}

@inproceedings{zhou2026fidelitygpt,
  title={FidelityGPT: Correcting Decompilation Distortions with Retrieval Augmented Generation},
  author={Zhou, Zhiping and Li, Xiaohong and Feng, Ruitao and Zhang, Yao and Li, Yuekang and Feng, Wenbu and Wang, Yunqian and Li, Yuqing},
  booktitle={Network and Distributed System Security (NDSS) Symposium 2026}
}

@article{chang2024survey,
  title={A survey on evaluation of large language models},
  author={Chang, Yupeng and Wang, Xu and Wang, Jindong and Wu, Yuan and Yang, Linyi and Zhu, Kaijie and Chen, Hao and Yi, Xiaoyuan and Wang, Cunxiang and Wang, Yidong and others},
  journal={ACM Transactions on Intelligent Systems and Technology},
  volume={15},
  number={3},
  pages={1--45},
  year={2024},
  publisher={ACM New York, NY}
}

@article{zhao2023survey,
  title={A survey of large language models},
  author={Zhao, Wayne Xin and Zhou, Kun and Li, Junyi and Tang, Tianyi and Wang, Xiaolei and Hou, Yupeng and Min, Yingqian and Zhang, Beichen and Zhang, Junjie and Dong, Zican and others},
  journal={arXiv preprint arXiv:2303.18223},
  year={2023}
}

@article{lee2023guess,
  title={Guess \& Sketch: Language Model Guided Transpilation},
  author={Lee, Celine and Mahmoud, Abdulrahman and Kurek, Michal and Campanoni, Simone and Brooks, David and Chong, Stephen and Wei, Gu-Yeon and Rush, Alexander M},
  journal={arXiv preprint arXiv:2309.14396},
  year={2023}
}

@article{cummins2024meta,
  title={Meta large language model compiler: Foundation models of compiler optimization},
  author={Cummins, Chris and Seeker, Volker and Grubisic, Dejan and Roziere, Baptiste and Gehring, Jonas and Synnaeve, Gabriel and Leather, Hugh},
  journal={arXiv preprint arXiv:2407.02524},
  year={2024}
}

@article{guo2024deepseek,
  title={DeepSeek-Coder: When the Large Language Model Meets Programming--The Rise of Code Intelligence},
  author={Guo, Daya and Zhu, Qihao and Yang, Dejian and Xie, Zhenda and Dong, Kai and Zhang, Wentao and Chen, Guanting and Bi, Xiao and Wu, Yu and Li, YK and others},
  journal={arXiv preprint arXiv:2401.14196},
  year={2024}
}

@inproceedings{10.1145/3597503.3639091,
author = {Li, Zongjie and Wang, Chaozheng and Ma, Pingchuan and Liu, Chaowei and Wang, Shuai and Wu, Daoyuan and Gao, Cuiyun and Liu, Yang},
title = {On Extracting Specialized Code Abilities from Large Language Models: A Feasibility Study},
year = {2024},
isbn = {9798400702174},
publisher = {Association for Computing Machinery},
address = {New York, NY, USA},
url = {https://doi.org/10.1145/3597503.3639091},
doi = {10.1145/3597503.3639091},
booktitle = {Proceedings of the IEEE/ACM 46th International Conference on Software Engineering},
articleno = {74},
numpages = {13},
location = {Lisbon, Portugal},
series = {ICSE '24}
}

@article{webson2021prompt,
  title={Do prompt-based models really understand the meaning of their prompts?},
  author={Webson, Albert and Pavlick, Ellie},
  journal={arXiv preprint arXiv:2109.01247},
  year={2021}
}

@inproceedings{ziegler2022productivity,
  title={Productivity assessment of neural code completion},
  author={Ziegler, Albert and Kalliamvakou, Eirini and Li, X Alice and Rice, Andrew and Rifkin, Devon and Simister, Shawn and Sittampalam, Ganesh and Aftandilian, Edward},
  booktitle={Proceedings of the 6th ACM SIGPLAN International Symposium on Machine Programming},
  pages={21--29},
  year={2022}
}

@article{zhang2023siren,
  title={Siren's song in the AI ocean: a survey on hallucination in large language models},
  author={Zhang, Yue and Li, Yafu and Cui, Leyang and Cai, Deng and Liu, Lemao and Fu, Tingchen and Huang, Xinting and Zhao, Enbo and Zhang, Yu and Chen, Yulong and others},
  journal={arXiv preprint arXiv:2309.01219},
  year={2023}
}

@article{huang2023survey,
  title={A survey on hallucination in large language models: Principles, taxonomy, challenges, and open questions},
  author={Huang, Lei and Yu, Weijiang and Ma, Weitao and Zhong, Weihong and Feng, Zhangyin and Wang, Haotian and Chen, Qianglong and Peng, Weihua and Feng, Xiaocheng and Qin, Bing and others},
  journal={ACM Transactions on Information Systems},
  year={2023},
  publisher={ACM New York, NY}
}

@inproceedings{cuconasu2024power,
  title={The power of noise: Redefining retrieval for rag systems},
  author={Cuconasu, Florin and Trappolini, Giovanni and Siciliano, Federico and Filice, Simone and Campagnano, Cesare and Maarek, Yoelle and Tonellotto, Nicola and Silvestri, Fabrizio},
  booktitle={Proceedings of the 47th International ACM SIGIR Conference on Research and Development in Information Retrieval},
  pages={719--729},
  year={2024}
}

@article{gao2023retrieval,
  title={Retrieval-augmented generation for large language models: A survey},
  author={Gao, Yunfan and Xiong, Yun and Gao, Xinyu and Jia, Kangxiang and Pan, Jinliu and Bi, Yuxi and Dai, Yixin and Sun, Jiawei and Wang, Haofen and Wang, Haofen},
  journal={arXiv preprint arXiv:2312.10997},
  volume={2},
  number={1},
  year={2023}
}

@article{wei2022chain,
  title={Chain-of-thought prompting elicits reasoning in large language models},
  author={Wei, Jason and Wang, Xuezhi and Schuurmans, Dale and Bosma, Maarten and Xia, Fei and Chi, Ed and Le, Quoc V and Zhou, Denny and others},
  journal={Advances in neural information processing systems},
  volume={35},
  pages={24824--24837},
  year={2022}
}

@misc{apple-silicon,
  title={Apple Silicon},
  howpublished = "\url{https://developer.apple.com/documentation/apple-silicon}",
  author={Apple},
 year = {2024},
}

@misc{rk3588,
  title={Useful resources for developing with the RK3588},
  howpublished = "\url{https://github.com/choushunn/awesome-RK3588/tree/develop}",
  year = {2024},
}

@misc{gpt-4,
  title={GPT-4 is OpenAI's most advanced system, producing safer and more useful responses},
  howpublished = "\url{https://openai.com/index/gpt-4/}",
  author={OpenAI},
 year = {2024},
}

@misc{apple-m4,
  title={Apple Introduces M4 Chip},
  howpublished = "\url{https://www.apple.com/ci/newsroom/2024/05/apple-introduces-m4-chip/}",
  author={Apple},
 year = {2024},
}

@misc{apple-porting,
  title={Porting Your macOS Apps to Apple Silicon},
  howpublished = "\url{https://developer.apple.com/documentation/apple-silicon/porting-your-macos-apps-to-apple-silicon}",
  author={Apple},
 year = {2024},
}

@misc{frumusanu2020amazon,
  title={Amazon's arm-based Graviton2 against AMD and Intel: Comparing cloud compute},
  author={Frumusanu, Andrei},
  year={2020}
}

@misc{shah2021we,
  title={We're closing the gap with Arm and x86, claims SiFive: New RISC-V CPU core for PCs, servers, mobile incoming {\'s} The Register},
  author={Shah, Agam},
  year={2021}
}

@misc{hisilicon,
  title={Kunpeng Chipsets},
  howpublished = "\url{https://www.hisilicon.com/en/products/Kunpeng/Huawei-Kunpeng/Huawei-Kunpeng-920}",
  author={HiSilicon},
 year = {2024},
}

@misc{qualcomm-snapdragon,
  title={Snapdragon X Elite},
  howpublished = "\url{https://www.qualcomm.com/products/mobile/snapdragon/laptops-and-tablets/snapdragon-x-elite}",
  author={Qualcomm},
 year = {2024},
}

@misc{windows-prism,
  title={How emulation works on Arm},
  howpublished = "\url{https://learn.microsoft.com/en-us/windows/arm/apps-on-arm-x86-emulation}",
  author={Microsoft},
 year = {2024},
}

@misc{apple-rosetta,
  title={Rosetta Translation Environment},
  howpublished = "\url{https://developer.apple.com/documentation/apple-silicon/about-the-rosetta-translation-environment}",
  author={Apple},
 year = {2024},
}

@misc{exagear,
  title={Huawei Kunpeng Exagear},
  howpublished = "\url{https://www.hikunpeng.com/document/detail/en/kunpengdevps/ug-exagear/usermanual/kunpengexagear_06_0003.html}",
  author={Huawei},
 year = {2024},
}

@misc{zydis,
  title={Fast and lightweight x86/x86-64 disassembler and code generation library},
  howpublished = "\url{https://github.com/zyantific/zydis}",
  author={zyantific},
 year = {2024},
}

@misc{yyjson,
  title={The fastest JSON library in C},
  howpublished = "\url{https://github.com/ibireme/yyjson}",
  author={ibireme},
 year = {2024},
}

@misc{c-algo,
  title={A library of common data structures and algorithms written in C},
  howpublished = "\url{https://github.com/fragglet/c-algorithms}",
  author={Simon Howard},
 year = {2024},
}

@article{henning2006spec,
  title={SPEC CPU2006 benchmark descriptions},
  author={Henning, John L},
  journal={ACM SIGARCH Computer Architecture News},
  volume={34},
  number={4},
  pages={1--17},
  year={2006},
  publisher={ACM New York, NY, USA}
}

@inproceedings{bucek2018spec,
  title={SPEC CPU2017: Next-generation compute benchmark},
  author={Bucek, James and Lange, Klaus-Dieter and v. Kistowski, J{\'o}akim},
  booktitle={Companion of the 2018 ACM/SPEC International Conference on Performance Engineering},
  pages={41--42},
  year={2018}
}

@article{kusswurm2014x87,
  title={X87 FPU Programming},
  author={Kusswurm, Daniel and Kusswurm, Daniel},
  journal={Modern X86 Assembly Language Programming: 32-bit, 64-bit, SSE, and AVX},
  pages={103--131},
  year={2014},
  publisher={Springer}
}

@misc{neon,
  title={What is NEON?},
  howpublished = "\url{https://developer.arm.com/documentation/dht0002/a/Introducing-NEON/What-is-NEON-}",
  author={ARM},
 year = {2024},
}

@inproceedings{lee2018efficient,
  title={Efficient SIMD implementation for accelerating convolutional neural network},
  author={Lee, Sung-Jin and Park, Sang-Soo and Chung, Ki-Seok},
  booktitle={Proceedings of the 4th International Conference on Communication and Information Processing},
  pages={174--179},
  year={2018}
}

@misc{qemu-emit,
  title={emit.c.inc},
  howpublished = "\url{https://github.com/qemu/qemu/blob/bc6afa1c711da5b4f37c9685a812c77b114d84cb/target/i386/tcg/emit.c.inc#L2468}",
  author={QEMU},
 year = {2024},
}

@inproceedings{bellard2005qemu,
  title={QEMU, a fast and portable dynamic translator.},
  author={Bellard, Fabrice},
  booktitle={USENIX annual technical conference, FREENIX Track},
  volume={41},
  number={46},
  pages={10--5555},
  year={2005},
  organization={California, USA}
}

@article{bartholomew2006qemu,
  title={Qemu: a multihost, multitarget emulator},
  author={Bartholomew, Daniel},
  journal={Linux Journal},
  volume={2006},
  number={145},
  pages={3},
  year={2006},
  publisher={Belltown Media Houston, TX}
}

@article{quynh2015unicorn,
  title={Unicorn: Next generation cpu emulator framework},
  author={Quynh, NGUYEN Anh and Vu, DANG Hoang},
  journal={BlackHat USA},
  volume={476},
  year={2015}
}

@article{li2024tiaozhuan,
  title={Tiaozhuan: A General and Efficient Indirect Branch Optimization for Binary Translation},
  author={Li, Xinyu and Guo, Guangyao and Lan, Yanzhi and Xue, Feng and Han, Chenji and Niu, Gen and Zhang, Fuxin},
  journal={ACM Transactions on Architecture and Code Optimization},
  year={2024},
  publisher={ACM New York, NY}
}

@article{xie2024instruction,
  title={An Instruction Inflation Analyzing Framework for Dynamic Binary Translators},
  author={Xie, Benyi and Yan, Yue and Yan, Chenghao and Tao, Sicheng and Zhang, Zhuangzhuang and Li, Xinyu and Lan, Yanzhi and Wu, Xiang and Liu, Tianyi and Zhang, Tingting and others},
  journal={ACM Transactions on Architecture and Code Optimization},
  volume={21},
  number={2},
  pages={1--25},
  year={2024},
  publisher={ACM New York, NY}
}

@inproceedings{wang2016general,
  title={A General Persistent Code Caching Framework for Dynamic Binary Translation ($\{$$\{$$\{$$\{$$\{$DBT$\}$$\}$$\}$$\}$$\}$)},
  author={Wang, Wenwen and Yew, Pen-Chung and Zhai, Antonia and McCamant, Stephen},
  booktitle={2016 USENIX Annual Technical Conference (USENIX ATC 16)},
  pages={591--603},
  year={2016}
}

@inproceedings{song2019unleashing,
  title={Unleashing the Power of Learning: An Enhanced $\{$Learning-Based$\}$ Approach for Dynamic Binary Translation},
  author={Song, Changheng and Wang, Wenwen and Yew, Pen-Chung and Zhai, Antonia and Zhang, Weihua},
  booktitle={2019 USENIX Annual Technical Conference (USENIX ATC 19)},
  pages={77--90},
  year={2019}
}

@article{wang2018enhancing,
  title={Enhancing cross-isa dbt through automatically learned translation rules},
  author={Wang, Wenwen and McCamant, Stephen and Zhai, Antonia and Yew, Pen-Chung},
  journal={ACM SIGPLAN Notices},
  volume={53},
  number={2},
  pages={84--97},
  year={2018},
  publisher={ACM New York, NY, USA}
}

@inproceedings{jiang2020more,
  title={More with less--deriving more translation rules with less training data for dbts using parameterization},
  author={Jiang, Jinhu and Dong, Rongchao and Zhou, Zhongjun and Song, Changheng and Wang, Wenwen and Yew, Pen-Chung and Zhang, Weihua},
  booktitle={2020 53rd Annual IEEE/ACM International Symposium on Microarchitecture (MICRO)},
  pages={415--426},
  year={2020},
  organization={IEEE}
}

@inproceedings{jiang2024system,
  title={A System-Level Dynamic Binary Translator Using Automatically-Learned Translation Rules},
  author={Jiang, Jinhu and Liang, Chaoyi and Dong, Rongchao and Yang, Zhaohui and Zhou, Zhongjun and Wang, Wenwen and Yew, Pen-Chung and Zhang, Weihua},
  booktitle={2024 IEEE/ACM International Symposium on Code Generation and Optimization (CGO)},
  pages={423--434},
  year={2024},
  organization={IEEE}
}

@inproceedings{wang2011coremu,
  title={COREMU: a scalable and portable parallel full-system emulator},
  author={Wang, Zhaoguo and Liu, Ran and Chen, Yufei and Wu, Xi and Chen, Haibo and Zhang, Weihua and Zang, Binyu},
  booktitle={Proceedings of the 16th ACM symposium on Principles and practice of parallel programming},
  pages={213--222},
  year={2011}
}

@inproceedings{d2017hypermambo,
  title={Hypermambo-x64: Using virtualization to support high-performance transparent binary translation},
  author={d'Antras, Amanieu and Gorgovan, Cosmin and Garside, Jim and Goodacre, John and Luj{\'a}n, Mikel},
  booktitle={Proceedings of the 13th ACM SIGPLAN/SIGOPS International Conference on Virtual Execution Environments},
  pages={228--241},
  year={2017}
}

@article{spink2020retargetable,
  title={A retargetable system-level DBT hypervisor},
  author={Spink, Tom and Wagstaff, Harry and Franke, Bj{\"o}rn},
  journal={ACM Transactions on Computer Systems (TOCS)},
  volume={36},
  number={4},
  pages={1--24},
  year={2020},
  publisher={ACM New York, NY, USA}
}

@inproceedings{engelke2021efficient,
  title={Efficient LLVM-based dynamic binary translation},
  author={Engelke, Alexis and Okwieka, Dominik and Schulz, Martin},
  booktitle={Proceedings of the 17th ACM SIGPLAN/SIGOPS International Conference on Virtual Execution Environments},
  pages={165--171},
  year={2021}
}

@inproceedings{cytron1989efficient,
  title={An efficient method of computing static single assignment form},
  author={Cytron, Ron and Ferrante, Jeanne and Rosen, Barry K and Wegman, Mark N and Zadeck, F Kenneth},
  booktitle={Proceedings of the 16th ACM SIGPLAN-SIGACT symposium on Principles of programming languages},
  pages={25--35},
  year={1989}
}

@book{van2007static,
  title={Static single assignment for decompilation},
  author={Van Emmerik, Michael James},
  year={2007},
  publisher={University of Queensland}
}

@article{cytron1991efficiently,
  title={Efficiently computing static single assignment form and the control dependence graph},
  author={Cytron, Ron and Ferrante, Jeanne and Rosen, Barry K and Wegman, Mark N and Zadeck, F Kenneth},
  journal={ACM Transactions on Programming Languages and Systems (TOPLAS)},
  volume={13},
  number={4},
  pages={451--490},
  year={1991},
  publisher={ACM New York, NY, USA}
}

@inproceedings{hara2008chstone,
  title={Chstone: A benchmark program suite for practical c-based high-level synthesis},
  author={Hara, Yuko and Tomiyama, Hiroyuki and Honda, Shinya and Takada, Hiroaki and Ishii, Katsuya},
  booktitle={2008 IEEE International Symposium on Circuits and Systems (ISCAS)},
  pages={1192--1195},
  year={2008},
  organization={IEEE}
}

@article{gal2012exploring,
  title={Exploring coremark a benchmark maximizing simplicity and efficacy},
  author={Gal-On, Shay and Levy, Markus},
  journal={The Embedded Microprocessor Benchmark Consortium},
  volume={6},
  number={23},
  pages={87},
  year={2012}
}

@misc{coremark,
  title={CoreMark-PRO},
  author={EEMBC},
  howpublished = "\url{https://www.eembc.org/coremark/}",
  year = {2025},
}

@misc{intel8086,
  title={Silicon reverse-engineering: the Intel 8086 processor's flag circuitry},
  author={Ken Shirriff},
  howpublished = "\url{https://www.righto.com/2023/02/silicon-reverse-engineering-intel-8086.html}",
  year = {2023},
}

@article{guide2011intel,
  title={Intel{\textregistered} 64 and ia-32 architectures software developer’s manual},
  author={Guide, Part},
  journal={Volume 3B: system programming guide, Part},
  volume={2},
  number={11},
  pages={0--40},
  year={2011}
}

@article{horspool1980approach,
  title={An approach to the problem of detranslation of computer programs},
  author={Horspool, R. Nigel and Marovac, Nenad},
  journal={The Computer Journal},
  volume={23},
  number={3},
  pages={223--229},
  year={1980},
  publisher={The British Computer Society}
}

@article{smith1985design,
  title={The design of divide and conquer algorithms},
  author={Smith, Douglas R},
  journal={Science of Computer Programming},
  volume={5},
  pages={37--58},
  year={1985},
  publisher={Elsevier}
}

@misc{aarch64-push,
  title={Using the Stack in AArch64: Implementing Push and Pop},
  author={Jacob Bramley},
  howpublished = "\url{https://community.arm.com/arm-community-blogs/b/architectures-and-processors-blog/posts/using-the-stack-in-aarch64-implementing-push-and-pop}",
  year = {2015},
}

@inproceedings{cobleigh2003learning,
  title={Learning assumptions for compositional verification},
  author={Cobleigh, Jamieson M and Giannakopoulou, Dimitra and P{\u{a}}s{\u{a}}reanu, Corina S},
  booktitle={International Conference on Tools and Algorithms for the Construction and Analysis of Systems},
  pages={331--346},
  year={2003},
  organization={Springer}
}

@misc{arm-call,
  title={Procedure Call Standard for the Arm Architecture},
  author={ARM},
  howpublished = "\url{https://github.com/ARM-software/abi-aa/blob/2982a9f3b512a5bfdc9e3fea5d3b298f9165c36b/aapcs32/aapcs32.rst#the-base-procedure-call-standard}",
  year = {2025},
}

@misc{arm-abi,
  title={Application Binary Interface for the Arm® Architecture – The Base Standard},
  author={ARM},
  howpublished = "\url{https://github.com/ARM-software/abi-aa/blob/2982a9f3b512a5bfdc9e3fea5d3b298f9165c36b/bsabi32/bsabi32.rst}",
  year = {2025},
}

@inproceedings{beck2023atomig,
  title={AtoMig: automatically migrating millions lines of code from TSO to WMM},
  author={Beck, Martin and Bhat, Koustubha and Stri{\v{c}}evi{\'c}, Lazar and Chen, Geng and Behrens, Diogo and Fu, Ming and Vafeiadis, Viktor and Chen, Haibo and H{\"a}rtig, Hermann},
  booktitle={Proceedings of the 28th ACM International Conference on Architectural Support for Programming Languages and Operating Systems, Volume 2},
  pages={61--73},
  year={2023}
}

\begin{appendix}
  \clearpage
\section{Appendix}
\label{sec:appendix}

\begin{table}[H]
	\centering
	\caption{Evaluated software.}
	\label{tab:description}
	
	\resizebox{0.97\linewidth}{!}{
	  \begin{tabular}{c|l}
		\hline
		Name & Description   \\
		\hline
		\multirow{2}{*}{\texttt{adpcm}}  & Adaptive differential pulse code modulation  \\
		                & decoder and encoder \\ 
    \texttt{gsm}    & Linear predictive coding analysis of audio signal  \\
    \texttt{mips}   & Simplified MIPS processor  \\
		\texttt{motion} & Motion vector decoding for MPEG-2  \\
		\texttt{aes}    & Advanced encryption standard  \\
		\texttt{sha1}   & Secure hash algorithm  \\
		\texttt{blowfish} & Symmetric-key block cipher  \\
		\texttt{jpeg}   & Image decompression  \\
		
		\texttt{cjpeg} & JPEG image compression  \\
		\texttt{sha256} & Secure hash algorithm  \\
		\texttt{linpack} & Gaussian elimination with partial pivoting  \\
		\texttt{nnet} & Neural net  \\
		\texttt{loops} & Kernels based on Livermore loops  \\
		\texttt{parser} & XML parsing  \\
		\texttt{radix2} & Radix-2 fast Fourier transform  \\
		\texttt{zlib} & ZIP compression  \\
		
		\texttt{bzip2} & File compression  \\
		\texttt{xz} & File compressions  \\
		\texttt{mcf} & Single-depot vehicle scheduling  \\
		\texttt{libquantum} & Physics / Quantum computing  \\
		\texttt{hmmer} & Gene sequence database searching  \\
		
		\texttt{zydis} & Disassembler library  \\
		\texttt{yyjson} & JSON library  \\
		\hline
	  \end{tabular}
	}
	
\end{table}

\subsection{Performance of Static Concise IR Lifters}
\label{subsec:appendix:lifter}

We evaluated three state-of-the-art static concise IR lifters, including Mctoll,
RetDec, and Rev.ng, on the same set of 23 software used in our evaluation. We
found that these lifters failed in all cases. Specifically, RetDec and Rev.ng
can produce lifted IR code for all cases, but the lifted IR code is either
non-compilable or functionally incorrect. RetDec extensively skips and leaves
unimplemented instructions in its output undefined, while Rev.ng translated
binaries crash immediately upon execution. On the other hand, Mctoll fails to
lift any of the cases, with error messages reported in \T~\ref{tab:mctoll}. 

\begin{table}[htp]
	\centering
	\caption{Mctoll's Error Messages.}
	\label{tab:mctoll}
	
	\resizebox{0.99\linewidth}{!}{
	  \begin{tabular}{c|l}
		\hline
		Test & Error Message   \\
		\hline
		\multirow{2}{*}{\texttt{adpcm}}  & Branch instruction terminating basic block   \\
     & computing jump table \\
    \texttt{gsm}    & Unhandled memory referencing instruction  \\
    \multirow{2}{*}{\texttt{mips}}  & Branch instruction terminating basic block   \\
     & computing jump table \\
		\multirow{2}{*}{\texttt{motion}}  & Unhandled PUSH instruction with non-AddrRegFrm   \\
     & source operand \\
		\multirow{2}{*}{\texttt{aes}} & Null reaching definition found during reaching   \\
     & definition fixup \\
		\texttt{sha1}   & Alignment is neither 0 nor a power of 2  \\
		\multirow{2}{*}{\texttt{blowfish}} & Null reaching definition found during reaching   \\
     & definition fixup \\
		\multirow{2}{*}{\texttt{jpeg}}  & Branch instruction terminating basic block   \\
     & computing jump table \\
    
		\texttt{cjpeg} & Unknown opcode  \\
		\texttt{sha256} & Unknown opcode  \\
		\texttt{linpack} & Unknown opcode  \\
		\texttt{nnet} & Unknown opcode  \\
		\texttt{loops} & Unknown opcode  \\
		\texttt{parser} & Unknown opcode  \\
		\texttt{radix2} & Unknown opcode  \\
		\texttt{zlib} & Unknown opcode  \\
		
		\texttt{bzip2} & cast<Ty>() argument of incompatible type!  \\
		\texttt{xz} & Unexpected block terminator found  \\
		\texttt{mcf} & Unknown opcode  \\
		\texttt{libquantum} & Unhandled SSE Packed Single Prefix encountered  \\
		\multirow{2}{*}{\texttt{hmmer}}  & Branch instruction terminating basic block   \\
     & computing jump table \\
    
		\texttt{zydis} & Unexpected number of successors of switch block  \\
		\texttt{yyjson} & Unexpected block terminator found  \\
		\hline
	  \end{tabular}
	}
	
\end{table}

\subsection{Binary-Level Data Structure Inconsistency}
\label{subsec:appendix:stat}

\begin{figure}[htp]
    \centering
    \includegraphics[width=0.98\linewidth]{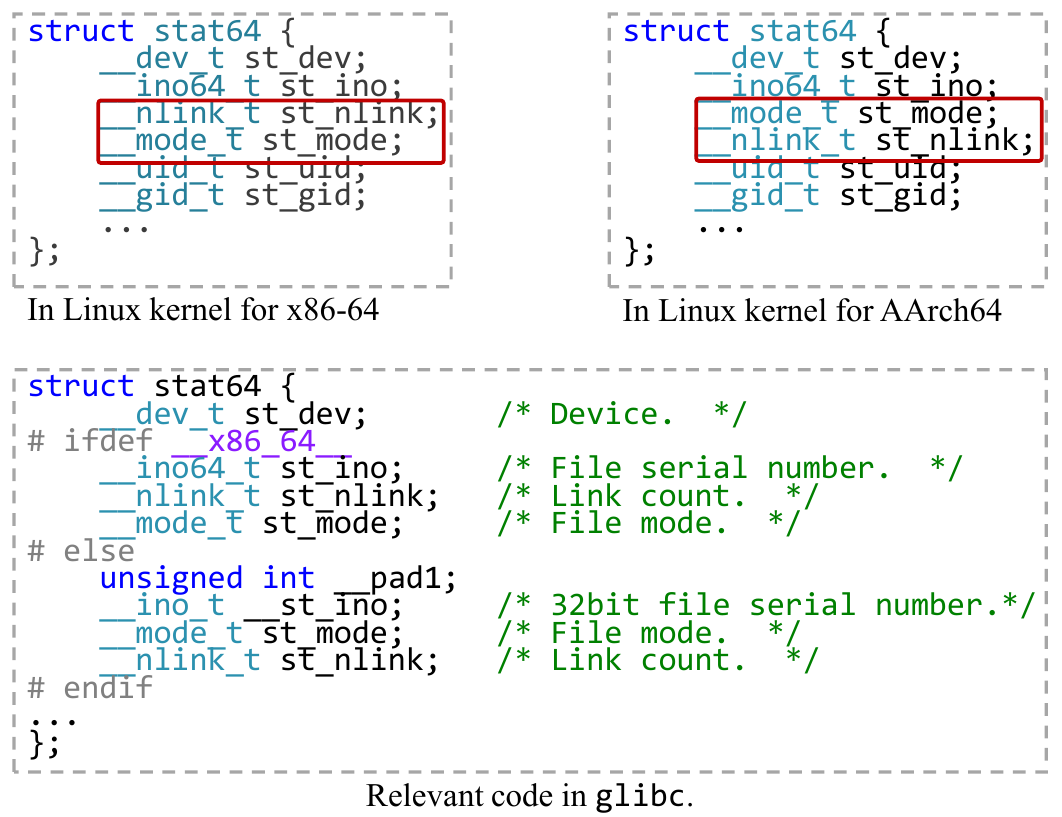}
	
    \caption{Binary-level data structure inconsistency.}
	
    \label{fig:stat}
\end{figure}

As shown in \F~\ref{fig:stat}, Linux kernel defines the \texttt{stat} structure
in different ways for
x86-64~\footnote{\raggedright \url{https://github.com/torvalds/linux/blob/8cf0b93919e13d1e8d4466eb4080a4c4d9d66d7b/arch/x86/include/uapi/asm/stat.h\#L42}}
and
AArch64~\footnote{\href{https://github.com/torvalds/linux/blob/8cf0b93919e13d1e8d4466eb4080a4c4d9d66d7b/arch/arm/include/uapi/asm/stat.h\#L57}{https://github.com/torvalds/linux/blob/8cf0b93919e13d1e8d4466eb-4080a4c4d9d66d7b/arch/arm/include/uapi/asm/stat.h\#L57}}
architectures. Specifically, the order of \texttt{st\_nlink} and
\texttt{st\_mode} are different. And \texttt{glibc} handles this difference by
explicitly parsing data returned by system calls according to the target
architecture~\footnote{\raggedright \url{https://github.com/lattera/glibc/blob/895ef79e04a953cac1493863bcae29ad85657ee1/sysdeps/unix/sysv/linux/x86/bits/stat.h\#L119}},
but still retains such binary-level data structure inconsistency. In this
example, \texttt{glibc} keep the same API for \texttt{fstat} system call.
Source-level references to specific members in \texttt{stat} structure are still
valid. However, since binary-level references are achieved with constant
offsets, the inconsistency in the layout of \texttt{stat} structure may cause
the translated binary to access wrong members. 

\end{appendix}

\end{document}